\documentclass[pre,twocolumn,amsmath,amssymb,nofootinbib,floatfix,superscriptaddress]{revtex4}

\usepackage{graphicx, bm, xcolor, enumerate, float}
\usepackage[normalem]{ulem}
\usepackage[breaklinks]{hyperref}

\makeatletter

\def\graphicscale{\twocolumn@sw{0.3}{0.4}}
\def\graphicthreescale{\twocolumn@sw{0.3}{0.4}}

\begin{document}

\title{Out-of-equilibrium finite-time percolation transitions and
  spinodal-like behaviors\\ after quenches across magnetic first-order
  transitions of Ising systems}

\author{Andrea Pelissetto}
\affiliation{Dipartimento di Fisica dell'Universit\`a di Roma ``La Sapienza"
  and INFN, Sezione di Roma, P.le Aldo Moro 5, I-00185 Roma, Italy}

\author{Davide Rossini} 
\affiliation{Dipartimento di Fisica dell'Universit\`a di Pisa and INFN,
  Largo Pontecorvo 3, I-56127 Pisa, Italy}

\author{Ettore Vicari} 
\affiliation{Dipartimento di Fisica dell'Universit\`a di Pisa,
  Largo Pontecorvo 3, I-56127 Pisa, Italy}

\date{\today}

\begin{abstract}
We analyze the out-of-equilibrium relaxational dynamics of
ferromagnetic Ising-like systems driven across their low-temperature
magnetic first-order transition (FOT) line, by sudden and slow
variations of an external homogenous magnetic field $h$.  As a
paradigmatic example, we consider a two-dimensional Ising system
driven across its low-temperature FOT line by suddently quenching $h$
from $h_i<0$ to $h>0$, inducing a transition from the negatively to
the positively magnetized phase.  We show the emergence of a dynamic
percolation transition at a finite critical time $t_c(h)$ along the
post-quench evolution for finite (sufficiently small) values of $h$,
which is marked by the percolation of the largest positive-spin
cluster and the antipercolation of the largest negative-spin cluster.
This out-of-equilibrium percolation transition displays a finite-size
scaling behavior as in the standard random-percolation case.  However,
while the fractal dimension of the percolating clusters is consistent
with the random-percolation value, the exponent controlling the
approach to criticality differs and depends on $h$.  We also show that
the percolation transition marks the passage from the metastable
negatively-magnetized phase to the stable positively-magnetized
one. Therefore, in the small-$h$ limit, the percolation critical
behavior is related to the spinodal-like behavior of the
magnetization, implying that the percolation time $t_c(h)$
exhibits a spinodal-like exponential dependence on $h$. The existence
of percolation transitions is likely a generic phenomenon at magnetic
Ising-like FOTs. For example, we observe an analogous behavior in
dynamic (Kibble-Zurek-like) protocols entailing slow variations of the
magnetic field driving the crossing of the FOT line.

\end{abstract}

\maketitle

\section{Introduction}
\label{intro}

Percolation transitions are notable phenomena appearing in different
contexts and fields, ranging from physics to chemistry, biology,
ecology, and social sciences~\cite{SA-book,Sahimi-book,Saberi-15}.
The percolation theory has been developed within the framework of
lattice models, or systems defined on graphs, in which
sites or bonds are occupied with a probability $p$.  As $p$ increases
from $p=0$, it reaches a threshold $p_c$ at which the occupied (or
activated) bonds form a percolating cluster, whose properties resemble
those of statistical systems undergoing continuous phase
transitions~\cite{Wilson-83,WK-74,Fisher-74,Wegner-76,PV-02}.  If the
activation probabilities of each site are independent, one recovers
the paradigmatic random percolation (RP) model, for which rigorous
exact results are available (see, e.g.,
Refs.~\cite{Kesten-book,Smirnov-01}).\footnote{RP models can be
mapped onto $q$-state Potts models in the formal limit $q\to
1$~\cite{Wu-82}, allowing us to reinterpret the geometric transition
as a standard one in a spin system.}  Percolation also emerges in
dynamic protocols when sites or bonds are reversibly added according
to a stochastic procedure~\cite{NZ-00}.  If the dynamic rules are
local, the system develops RP transitions as well. Different types of
percolation phenomena occur when considering nonreversible transition
rules (not satisfying detailed balance), such as directed percolation
(see, e.g., Refs.~\cite{Hinrichsen-00,Odor-04,AGKSZ-14}).

Other, apparently distinct, questions in statistical physics concern
the out-of-equilibrium dynamic behavior of systems driven across
classical and quantum first-order transitions (FOTs); they include
hysteresis, coarsening, and the emergence of dynamic spinodal-like
phase changes (see, e.g.,
Refs.~\cite{Binder-76,Binder-87,Bray-94,RTMS-94,RV-21,PV-24,CA-99,MM-00,LFGC-09,NIW-11,EH-13,PV-15,PV-16,PV-17,PV-17b,LZ-17,PPV-18,SW-18,PRV-18,Bar-etal-18,PRV-18-def,Fontana-19,LZW-19,PRV-20,DRV-20,CCP-21,SCD-21,CCEMP-22,TV-22,TS-23,PRV-25,PV-26,PRV-26,PRV-26-2,PRV-26-3}).
In this context, droplet nucleation~\cite{Binder-87,RTMS-94,PRV-26-3}
is usually the relevant mechanism providing a qualitative description
of the dynamics (see, e.g., Refs.~\cite{PV-17,PV-26,PRV-26,PRV-26-3}).
In the mean-field theory of FOTs~\cite{Binder-87}, the free-energy
density near the transition is characterized by two distinct
minima. They represent a stable and a metastable state, separated by a
free-energy barrier that diverges as the volume $V$, for $V\to
\infty$.  From a dynamical perspective, this implies that the system
can remain trapped in the metastable state for a duration that
diverges as $V \to \infty$. Within this framework, the spinodal line
is defined as the boundary where the metastable state becomes
unstable, i.e., the point at which the local free-energy minimum
corresponding to the metastable state disappears. In short-range
models there are no thermodynamic metastable states~\cite{Binder-87}.
Nevertheless, metastability is a ubiquitous feature of the dynamics of
systems close to FOTs: for instance, it emerges naturally when the
evolution drives the system across a FOT.  This is due to the fact
that equilibration requires the system to go across regions of
atypical configurations, i.e., to overcome free-energy barriers. In
short-range systems, these barriers are typically finite even in the
infinite-volume limit, at variance with the metastable states found in
the mean-field approach. Consequently, even as $V\to\infty$, the
dynamic metastable state decays in a finite time, depending
nontrivially on the external parameters which characterize the system.

In this paper we show that both percolative and spinodal-like
phenomena characterize the out-of-equilibrium dynamics of
ferromagnetic systems driven across FOTs, by sudden and slow variations
of the external magnetic field. We provide a geometric interpretation
of the passage from the metastable to the stable state across FOTs, in
terms of percolating clusters, identifying the dependence of the time
scale of the phase change on the external driving parameter.  We show
that out-of-equilibrium finite-time percolation transitions appear
within dynamic, local, and reversible physical protocols. Moreover,
we relate such finite-time percolation phenomena to the spinodal-like
behavior already observed along the out-of-equilibrium relaxational
dynamics across various FOTs of short-range spin
systems~\cite{PV-17,PRV-25,PV-26,PRV-26}.

We focus on the paradigmatic two-dimensional (2D) Ising model driven
across its low-temperature FOT line at zero magnetic field. In
particular, we analyze the out-of-equilibrium relaxational dynamics
following a quench of the external magnetic field $h$ at fixed
temperature. Starting from negatively magnetized configurations,
corresponding to equilibrium states for $h<0$, the system evolves in
the presence of a positive fixed $h>0$, thus effectively crossing the
magnetic FOT line. We show that the post-quench evolution develops an
out-of-equilibrium transition at a finite critical time $t_c$ after
the quench, marked by the percolation of the largest cluster of the
positive phase and associated with the transition from the metastable
negatively magnetized phase to the stable positively magnetized one.
We also present heuristic scaling arguments which allow us to obtain
quantitative predictions for the relevant time scales governing the
phase change, leading to a characteristic exponential dependence of
the time scale on the external field. We refer to such behavior as
spinodal-like, somehow corresponding to the spinodal of the mean-field
approach. These scenarios are verified by various numerical analyses
in nearest-neighbor square-lattice Ising models.~\footnote{Some
preliminary results were already presented in Ref.~\cite{PRV-26-ar}.}

The paper is organized as follows. In Sec.~\ref{sec2} we define the 2D
Ising model and the relaxational dynamics, which we consider to study
the out-of-equilibrium behavior across the FOTs.  In Sec.~\ref{quepro}
we outline the quench protocol and define the observables used
to monitor the dynamics.  Secs.~\ref{perctra} and~\ref{infmag} report
our numerical analyses for the quench dynamics in 2D Ising systems
with periodic boundary conditions (PBC). In Sec.~\ref{perctra} we report
results for the positive- and negative-spin clusters, showing the
emergence of a finite-time dynamic percolation transition at fixed
(sufficiently small) $h$ in the large-volume limit. In
Sec.~\ref{infmag} we analyze the post-quench dynamics of the
magnetization in the thermodynamic limit, we present some heuristic
arguments which allow us to predict the scaling behavior of the
magnetization in the limit $h\to 0$, showing that its spinodal-like
behavior in the small-$h$ limit is strictly related to the percolative
critical behavior and its dependence on $h$. In Sec.~\ref{sec4} we
show that the infinite-volume behavior of the magnetization is
independent of the boundary conditions, extending the analysis to
systems with fixed boundary conditions (FBC).  In Sec.~\ref{slowdyn} we
consider a different protocol, in which the system is driven across
the FOT by a slowly varying magnetic field (analogous to the standard
Kibble-Zurek protocols across continuous transitions), reporting
results that are qualitatively similar to those discussed in the
sudden quench case. Finally, in Sec.~\ref{conclu} we summarize and
draw our conclusions.

\section{The Ising model and its dynamics} \label{sec2}

We consider the paradigmatic square-lattice nearest-neighbor
Ising model with Hamiltonian
\begin{equation}
  H = - J \sum_{{\bm x},\mu} s_{\bm x} \, s_{{\bm x}+\hat{\mu}} - h
  \sum_{\bm x} s_{\bm x},\quad s_{\bm x}=\pm 1,
  \label{isiham}
\end{equation}
where the two sums go over the lattice bonds and sites,
respectively.  We focus on a ferromagnetic system and set $J=1$,
without loss of generality.  The partition function is
\begin{equation}
  Z(\beta,h) = \sum_{ \{ s_{\bm x} \} } \exp(-\beta H),
  \qquad \beta=1/T,
  \label{partfunc}
\end{equation}
where $\beta$ is the inverse temperature.  In our numerical study we
mostly consider systems of size $L\times L$ with PBC.
In Sec.~\ref{sec4}, we extend the analysis to systems with FBC,
setting $s_{\bm x} = 1$ on the boundary of the lattice.

The 2D Ising model presents a FOT line for $h=0$
and~\cite{Onsager-44,FF-69}
\begin{equation}
  \beta > \beta_c = {1\over 2} \ln (1 + \sqrt{2}).
  \label{fotline}
\end{equation}
Along this line, the infinite-volume magnetization
\begin{equation}
  M = {1\over V} \langle \, \sum_{\bm x} s_{\bm x} \, \rangle, \qquad V=L^2,
\end{equation}
is discontinuous. Indeed, 
\begin{equation}
  \lim_{h\to 0^\pm} M = \pm M_0,\qquad
  M_0=[1-(\sinh 2\beta)^{-4}]^{1/8},
  \label{m02d}
\end{equation}
see, e.g., Ref.~\cite{ID-book}.

Our numerical study focuses on the out-of-equilibrium relaxational
dynamic behavior of the system when it is driven across the FOT line.
For this purpose, we consider a single-spin heat-bath Monte Carlo (MC)
dynamics, which provides a standard realization of a purely
relaxational dynamics~\cite{Binder-76}. At each site, the spin is
updated by sampling from the conditional probability distribution $P
\sim e^{-\beta H}$, with the neighboring spins held fixed. The update
follows a checkerboard scheme: All spins on even sites are first
updated, followed by all spins on odd sites.  One time unit
corresponds to a complete lattice sweep. For fixed $\beta$ and $h$,
the resulting MC dynamics is not strictly reversible, due to the
sequential update of the two sublattices.  This also implies that the
time variable $t$ can only assume integer values.\footnote{A
reversible MC dynamics could be implemented by randomly selecting
spins at each update step. This alternative choice, giving rise to a
slower dynamics (by approximately a factor four at
criticality~\cite{PV-17b}), would also allow for noninteger values of
the time variable.  Indeed, defining the time unit as $L^2$ spin
updates, to ensure a well-defined infinite-volume limit at fixed $t$,
one might consider time steps of order $1/L^2$. In the $L\to \infty$
limit, this naturally leads to a continuous-time evolution.  However,
we expect that this alternative MC dynamics would lead to analogous
results, when crossing the magnetic FOT line.}

It is important to stress that any purely relaxational dynamics
is expected to provide the same qualitative behavior.  Thus, one may
alternatively use a Metropolis dynamics~\cite{Metropolis:1953am}, for
which the time evolution is obtained by local spin flips with
probability
\begin{equation}
  P(s_{{\bm x}}\to -s_{{\bm x}}) = {\rm Min}(1,e^{-\Delta H}),
  \label{metroup}
\end{equation}
where $\Delta H$ is the change of the Hamiltonian when replacing
$s_{{\bm x}}$ with $ -s_{{\bm x}}$.  The Metropolis and heat-bath
dynamics are both examples of a relaxational dynamics without
conservation laws, thus the results for the two cases should be
qualitatively similar.  Quantitatively, however, the Metropolis
dynamics is faster than the heat-bath one, since in Ising systems the
heat-bath dynamics is also a Metropolis-Hastings dynamics, but with a
smaller acceptance rate~\cite{Hastings-70,PV-26}.

\section{The quench protocol}
\label{quepro}

The quenching protocol starts at $t=0$ from equilibrated
negatively magnetized configurations at $\beta>\beta_c$. They are
obtained by performing heat-bath MC simulations at $h_i=0$, starting
from the fully ordered configuration $s_{\bm x} = -1$ for all $\bm x$.
We consider system sizes so large that a spontaneous
transition from the negative phase to the opposite one never
occurs.\footnote{Indeed, the typical time required for such a
transition scales as $L^2 \exp(2 \beta \kappa L)$, where $\kappa$ is
the planar interface tension (see, e.g., Ref.~\cite{PV-17b}), up to a
prefactor of order one. For $\beta=1.2\beta_c$, it scales as $L^2
\exp(0.666 L)$, corresponding to a time scale of order $10^{68}$ for
$L=200$.  Since all simulations are performed for $L\ge 200$, such
transitions never occur in practice.}  Note that starting from
states equilibrated at finite negative values of $h$ would not change
the general picture and would only lead to a trivial time shift.

The post-quench evolution, in the presence of a positive fixed
field $h>0$, is monitored by the time-dependent magnetization
\begin{equation}
  M(t) = {1\over V} \, \big\langle \sum_{\bm x} s_{\bm x} \big\rangle_t,
  \label{magn2}
\end{equation}
where the average is taken over both the initial equilibrium
configurations and the post-quench stochastic trajectories at fixed time $t$.
The percolative nature of the dynamic transition is probed by looking
at clusters of equally oriented spins. For each configuration, clusters
with positive (negative) magnetization are defined as the connected
components of the subset ${\cal B}_+$ (${\cal B}_-$) of lattice links
$\langle {\bm x}{\bm y}\rangle$ such that $s_{\bm x} = s_{\bm y} = 1$
($s_{\bm x} = s_{\bm y} = -1$).  These are identified using the
Hoshen-Kopelman algorithm~\cite{HK-76}. The size of each cluster is
defined as the number of sites belonging to it. We focus
on the fixed-time averages of the sizes $S_+$ and $S_-$ of the largest
clusters with positive and negative magnetization, respectively,
\begin{equation}
  \Sigma_\pm(t) = {\langle S_\pm \rangle_t\over V}, \qquad R_s(t)
  = {\langle S_- \rangle_t \over \langle S_+ \rangle_t}.
  \label{drddef}
\end{equation}
We also determine the sizes $S_{n,\pm}$ of the $n$th largest clusters
for a few values of $n$ and the corresponding quantities
\begin{equation}
  \Sigma_{n,\pm}(t) = {\langle S_{n,\pm} \rangle_t\over V}.
\end{equation}

In the following sections we report numerical results at fixed
$\beta = 1.2\,\beta_c$ and for several values of $h>0$. We do not observe
qualitative changes for other values of $\beta$, as long as
$\beta>\beta_c$.  For each $h$, MC simulations are performed for sizes
up to $L = 4000$, generating a large number of independent
trajectories, ranging from 2000 (for the largest systems) to 50000.

\section{Dynamic percolation transitions}
\label{perctra}

In this section we study the dynamic behavior of the model at fixed
post-quench magnetic field $h$, showing that, along the
out-of-equilibrium evolution, the system undergoes a geometric
percolation transition of the stable-phase clusters at a finite time
$t_c$ after quenching, which depends on $h$.

%%%%%%%%%%%%%%%%%%%%%%%%%%%%%%%%%%%%%%%%%%%%%%%%%%%%%%%%%%%%%%%%%%%%%%%
\begin{figure}[tbp]
  \includegraphics[width=0.9\columnwidth]{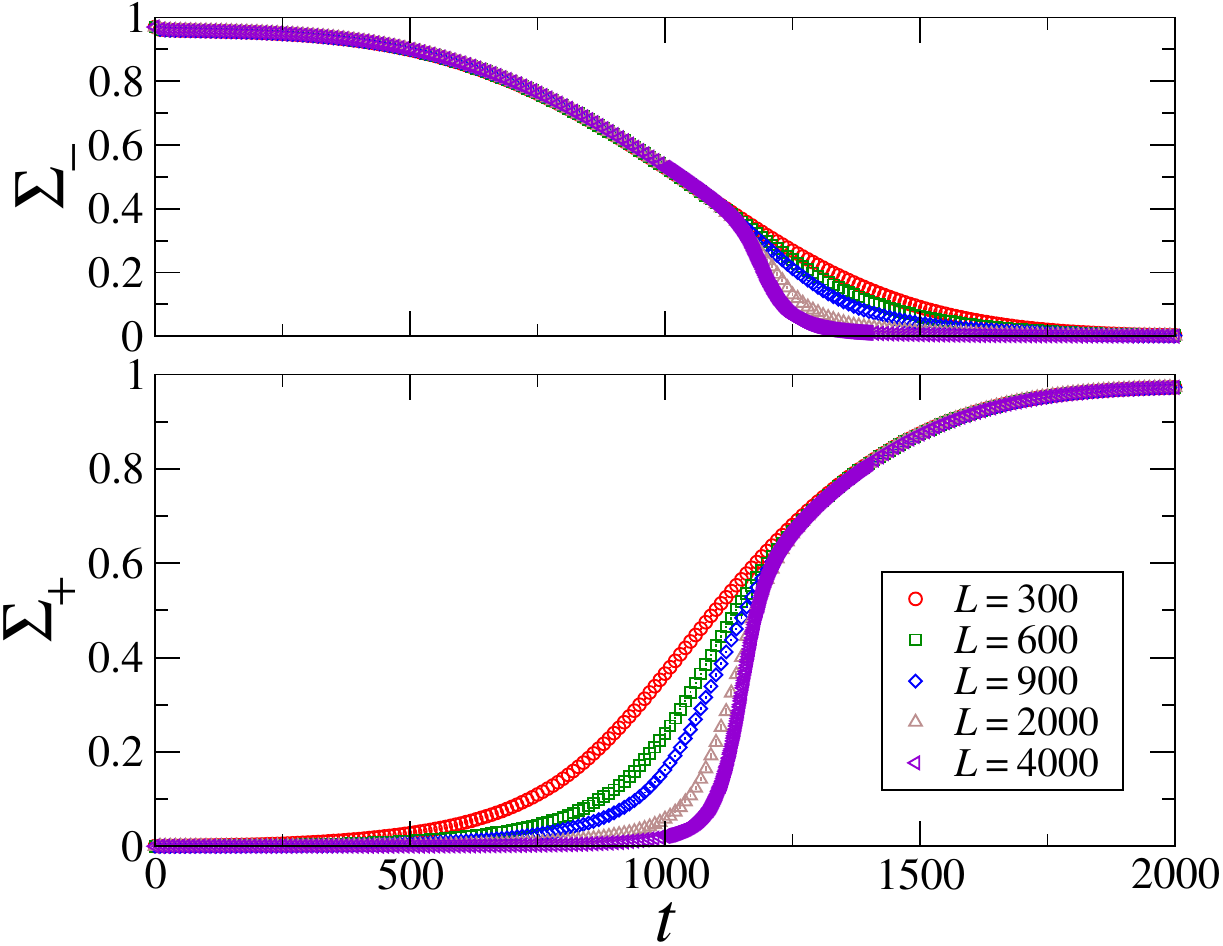}
  \caption{The rescaled average sizes $\Sigma_+$ (bottom) and
    $\Sigma_-$ (top) of the largest clusters with positive and negative
    magnetization, defined in Eq.~\eqref{drddef}, as functions of the
    post-quench time $t$, for $h=0.07$ and several sizes $L$.
    Statistical errors are not visible, as they are smaller than
    the symbol size.  }
\label{cluster2D}
\end{figure}
%%%%%%%%%%%%%%%%%%%%%%%%%%%%%%%%%%%%%%%%%%%%%%%%%%%%%%%%%%%%%%%%%%%%%%%

%%%%%%%%%%%%%%%%%%%%%%%%%%%%%%%%%%%%%%%%%%%%%%%%%%%%%%%%%%%%%%%%%%%%%%%
\begin{figure}[!b]
  \includegraphics[width=0.98\columnwidth]{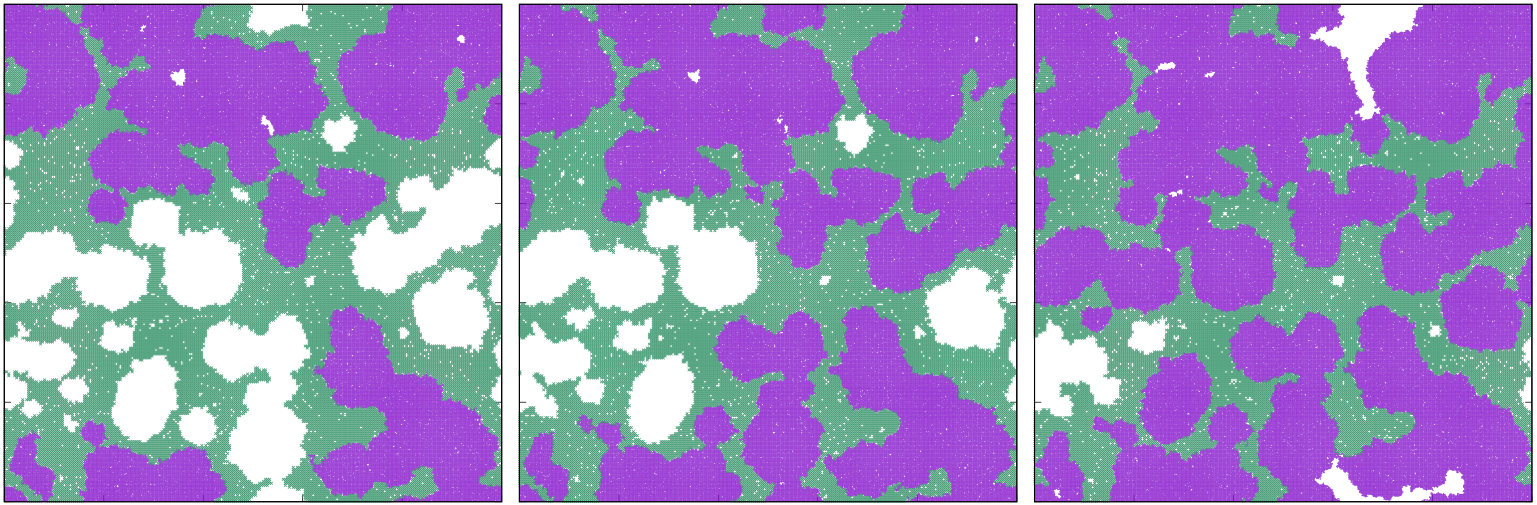}
  \caption{Snapshots of the configurations for $h = 0.07$ and
    $L=1000$, at times $t = 1144$ (left), $t = t_c = 1174$ (center),
    and $t = 1204$ (right). Violet and green sites correspond to the
    largest clusters of positive and negative spins, respectively,
    while white regions denote smaller (both positive and negative)
    clusters. The size of the largest positive cluster is $368001$ at
    $t = 1144$, $498264$ at $t=t_c=1174$, and $637204$ at
    $t=1204$. Its growth is primarily due to the merging with smaller
    clusters, since the boundaries of the clusters are essentially
    unchanged.}
\label{clusterfig2D}
\end{figure}
%%%%%%%%%%%%%%%%%%%%%%%%%%%%%%%%%%%%%%%%%%%%%%%%%%%%%%%%%%%%%%%%%%%%%%%

Figure~\ref{cluster2D} shows results for the ratios $\Sigma_\pm(t)$
defined in Eq.~\eqref{drddef}, for $h = 0.07$ and several values of
$L$. We identify two distinct small-time and large-time regimes.  In
the small-time regime the volume fraction of the positively magnetized
clusters vanishes in the infinite-volume limit, while a fraction of
the volume belongs to a single negatively magnetized percolating
cluster. In the large-time regime the opposite behavior is observed,
with a single positively magnetized percolating cluster coexisting
with small negatively magnetized domains.  These two regimes are
separated by a narrow region, around $t= t_c\approx 1200$,
characterized by the simultaneous sharp variation of both
$\Sigma_+(t)$ and $\Sigma_-(t)$, which becomes sharper and sharper
with increasing $L$.  As $t$ increases across $t_c$, positive-spin
clusters rapidly aggregate, forming a single macroscopic cluster,
while large negative-spin clusters
disappear. Figure~\ref{clusterfig2D} provides a sketch of the typical
rapid evolution of the configurations in the vicinity of the
transition time $t_c$ for a system of size $L=1000$.  Over a short
time window of 60 iterations, the size of the largest positive cluster
increases significantly ($\Sigma_+$ approximately takes the values
0.37, 0.50, and 0.64, as time increases). Note that the growth is
mainly driven by merging processes---different positive clusters
aggregate---and not by the growth of the single clusters, i.e.,
cluster boundaries are essentially unchanged in this short time
window.

Since the sharp change of $\Sigma_+$ and $\Sigma_-$ occurs in the same
range of values of $t$, it is natural to conjecture that the
percolation of positive clusters and the inverse percolation of
negative clusters occur at the same finite time $t_c$.  Under this
assumption, $\Sigma_\pm(t)$ is expected to exhibit the typical
finite-size scaling (FSS) behavior at percolation
transitions~\cite{SA-book,Sahimi-book,Saberi-15},
\begin{equation}
  \Sigma_\pm(t,L) \approx L^{-\delta_\pm} {\cal F}_\pm(X), \qquad
  X = (t-t_c) L^w,
  \label{Dpiu-scaling}
\end{equation}
where
\begin{equation}
  \delta_\pm \equiv d-d_\pm = 2-d_\pm,
  \label{deltapm}
\end{equation}
with $d_\pm$ denoting the
critical fractal dimension of the clusters, and the exponent $w$
controlling the approach to criticality.

%%%%%%%%%%%%%%%%%%%%%%%%%%%%%%%%%%%%%%%%%%%%%%%%%%%%%%%%%%%%%%%%%%%%%%%
\begin{figure}[tbp]
  \includegraphics[width=0.9\columnwidth]{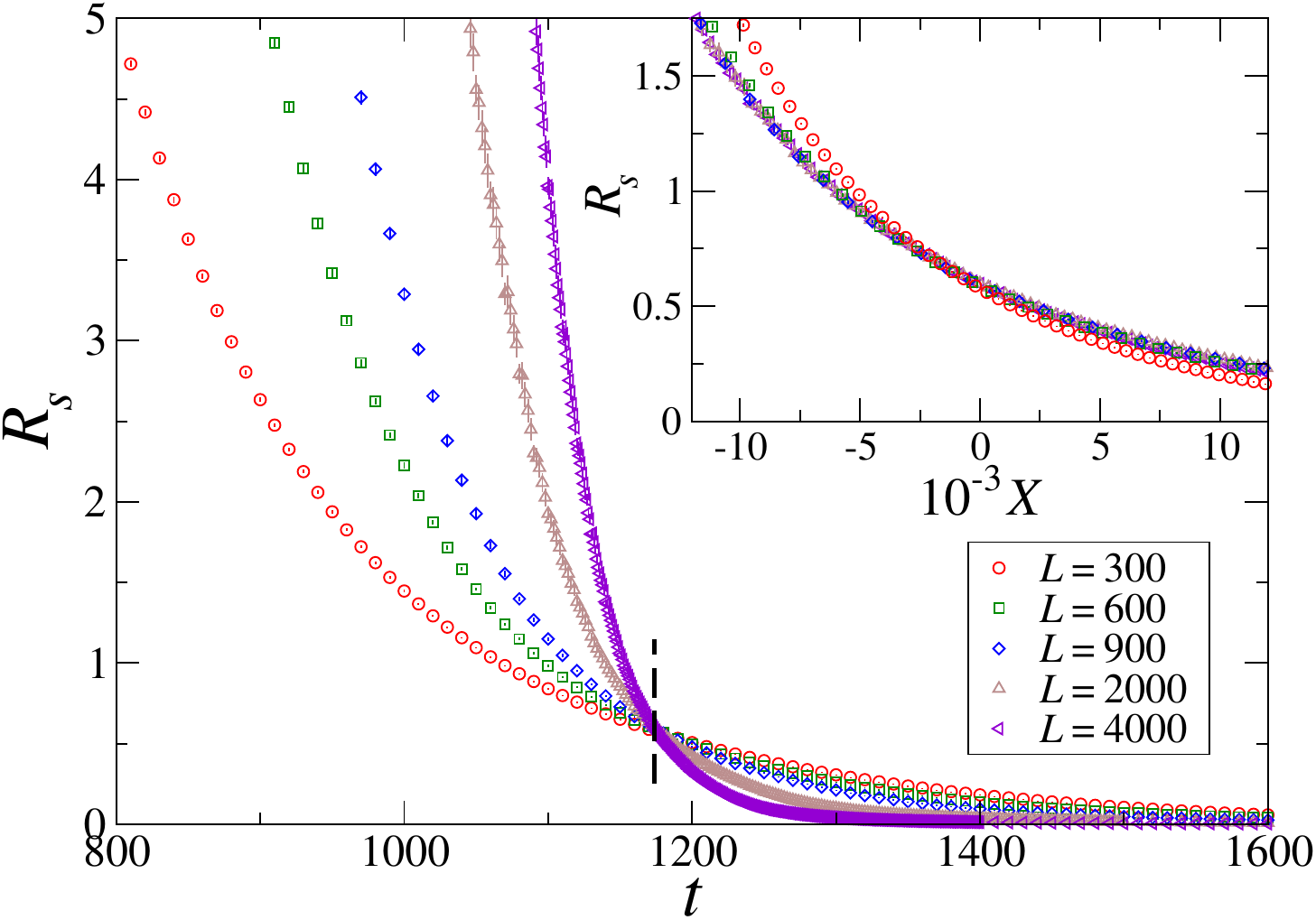}
  \caption{Ratio $R_s$ of the negative and positive largest cluster
    sizes vs $t$, for $h = 0.07$ (statistical errors are hardly
    visible). The vertical dashed line indicates the critical time
    $t_c = 1174$.  The inset shows $R_s$ vs $X = (t-t_c) L^{w}$, with
    $w = 0.68$.  The observed collapse of the curves for $L \gtrsim
    600$ supports the Ansatz~\eqref{Ansatz-FSS}, confirming the
    equality $d_+=d_-$ of the fractal dimensions of the positive and
    negative clusters.  }
\label{clusterratioh0p07}
\end{figure}
%%%%%%%%%%%%%%%%%%%%%%%%%%%%%%%%%%%%%%%%%%%%%%%%%%%%%%%%%%%%%%%%%%%%%%%

To verify whether the critical time is the same for positive and
negative clusters, we analyze the ratio $R_s$ of the negative and
positive largest cluster sizes, defined in Eq.~\eqref{drddef}. As
shown in Fig.~\ref{clusterratioh0p07}, curves for different system
sizes cross at a finite time, which can be identified with the
critical time $t_c$.  This behavior resembles that of the Binder
parameter in standard spin models~\cite{PV-02}. It confirms the
presence of a single transition time. Moreover it indicates that $d_+
= d_-$.  It is thus natural to conjecture the scaling form
\begin{equation}
  R_s(t,L) \equiv {\Sigma_-(t,L)\over \Sigma_+(t,L)} \approx
  {\cal F}_R(X).
  \label{Ansatz-FSS}
\end{equation}
Straightforward fits to the Ansatz~\eqref{Ansatz-FSS}, using low-order
polynomial interpolations of ${\cal F_R}(X)$, yield the estimates
\begin{equation}
  t_c = 1174(1),\quad w = 0.680(15), \quad {\rm for}\;\;h=0.07,
  \label{tcwest}
\end{equation}
where the errors take into account how $t_c$ and $w$ vary when
changing the order of the polynomial interpolation and discarding data
for $L<L_{\rm min}$ with increasing $L_{\rm min}$ (to somehow control
the scaling corrections).  The inset of Fig.~\ref{clusterratioh0p07}
demonstrates the quality of the data collapse obtained using these
estimates.

\begin{figure}[tbp]
  \includegraphics[width=0.9\columnwidth]{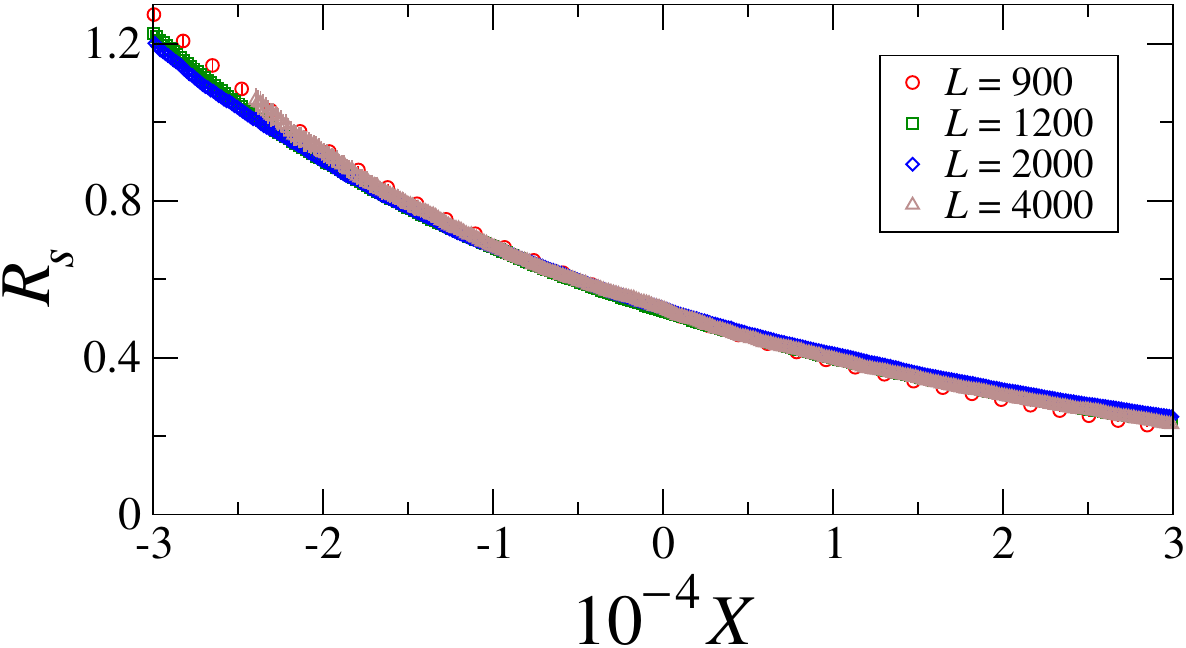}
  \caption{The ratio $R_s$ vs $X = (t-t_c) L^{w}$ for $h = 0.045$,
    with $t_c = 5921$ and $w = 0.52$.}
  \label{cluster2Dscalingh0p045}
\end{figure}

The same analysis has been performed for other values of
$h$, obtaining a good data collapse in all cases. See, for example,
Fig.~\ref{cluster2Dscalingh0p045} for the scaling plot of $R_s$ for
$h=0.045$.  The corresponding estimates of $t_c$ and $w$ are reported
in Table~\ref{table-td-2D}.  We note that the exponent $w$ depends on
$h$ and decreases as $h$ is reduced, suggesting that it vanishes in
the limit $h\to 0$. Consequently, the time window $\Delta t$ over
which the system undergoes the phase change, which scales as $L^{-w}$,
broadens as $h \to 0$.

%%%%%%%%%%%%%%%%%%%%%%%%%%%%%%%%%%%%%%%%%%%%%%%%%%%%%%%%%%%%%%%%%%%%%%%
%%%%%%%%%%%%%%%%%%%%%%%%%%%%%%%%%%%%%%%%%%%%%%%%%%%%%%%%%%%%%%%%%%%%%%%

%%%%%%%%%%%%%%%%%%%%%%%%%%%%%%%%%%%%%%%%%%%%%%%%%%%%%%%%%%%%%%%%%%%%%%%
\begin{table}[!b]
  \caption{The estimates of $t_c$, $w$, $\delta$, $\sigma_c = (\ln
    t_c)^2 h$ and $\zeta_c = (\sigma_c - \sigma_*) h^{-\theta}$ [using
      $\theta = 0.64(4)$ and $\sigma_* = 3.03(3)$].  The errors also
    account for the small variations of the results obtained when
    varying the minimum system size allowed in the fit, the range of
    data considered, and the degree of the polynomial used for the 
    scaling functions.}
  \label{table-td-2D}
  \vspace*{2mm}
  \begin{tabular}{lccccc}
    \hline\hline $h$ & $t_c$ & $w$ & $\delta$ &
    $\sigma_c$ & $\zeta_c$
    \\ \hline
    0.07 & 1174(1) & 0.680(15) & 0.106(4) & 3.497(1) &2.56(4) \\
    0.06 & 1987(3) & 0.64(2)   & 0.106(10)& 3.460(1) &2.61(5) \\
    0.05 & 3904(3) & 0.56(2)   & 0.094(6) & 3.419(1) &2.65(6) \\
    0.045& 5921(3) & 0.52(2)   & 0.089(9) & 3.395(1) &2.66(7)
    \\ \hline\hline
  \end{tabular}
\end{table}
%%%%%%%%%%%%%%%%%%%%%%%%%%%%%%%%%%%%%%%%%%%%%%%%%%%%%%%%%%%%%%%%%%%%%%%

This behavior is consistent with the progressive slowing down of the
dynamics with decreasing $h$.  To compare with the RP case, note that
the probability that a bond is occupied, i.e., that it belongs to a
positive-spin cluster, should be directly related to the
time-dependent magnetization $M(t)$, which is regular at $t_c$, as
discussed in the next section. This allows us to relate the deviation
from $t_c$ to the devation of the probability from its critical value
$p_c$ of RP, i.e., $t - t_c \sim p - p_c$. Therefore, if the
transition were of RP type, we would expect $w = w_{\rm RP} =
3/4$~\cite{Kesten-book,Smirnov-01} for all values of $h$.  This
relation is clearly excluded by the numerical data. The difference
between the estimated $w$ and $w_{\rm RP}$ is too large (about ten
error bars for $h = 0.045$) to be plausibly interpreted as due to
scaling corrections.

To verify whether the value of $R_s$ at the observed percolative
transitions is universal, i.e., independent of $h$, we determine
$R_s(t_c)$ at the various values of $h$. We find $0.60(1)$, $0.56(2)$,
$0.53(2)$, $0.53(2)$ for $h = 0.07$, $0.06$, $0.05$, $0.045$.  These
results show that for $t=t_c$ the positive-spin largest cluster is
approximately twice the size of its negative-spin counterpart.
However, they are apparently $h$ dependent with a finite limit for
$h\to 0$.\footnote{Note that the variable $R_s(t)$ does not have a
natural counterpart in RP, although on the square lattice bond
percolation occurs for $p = p_c =1/2$, the percolation of occupied
bonds occurs together with the antipercolation of empty bonds.
However, because of the exact symmetry, the analogous ratio $R_s(t_c)$
is one at the transition.}

%%%%%%%%%%%%%%%%%%%%%%%%%%%%%%%%%%%%%%%%%%%%%%%%%%%%%%%%%%%%%%%%%%%%%%%
\begin{figure}[!t]
  \includegraphics[width=0.9\columnwidth]{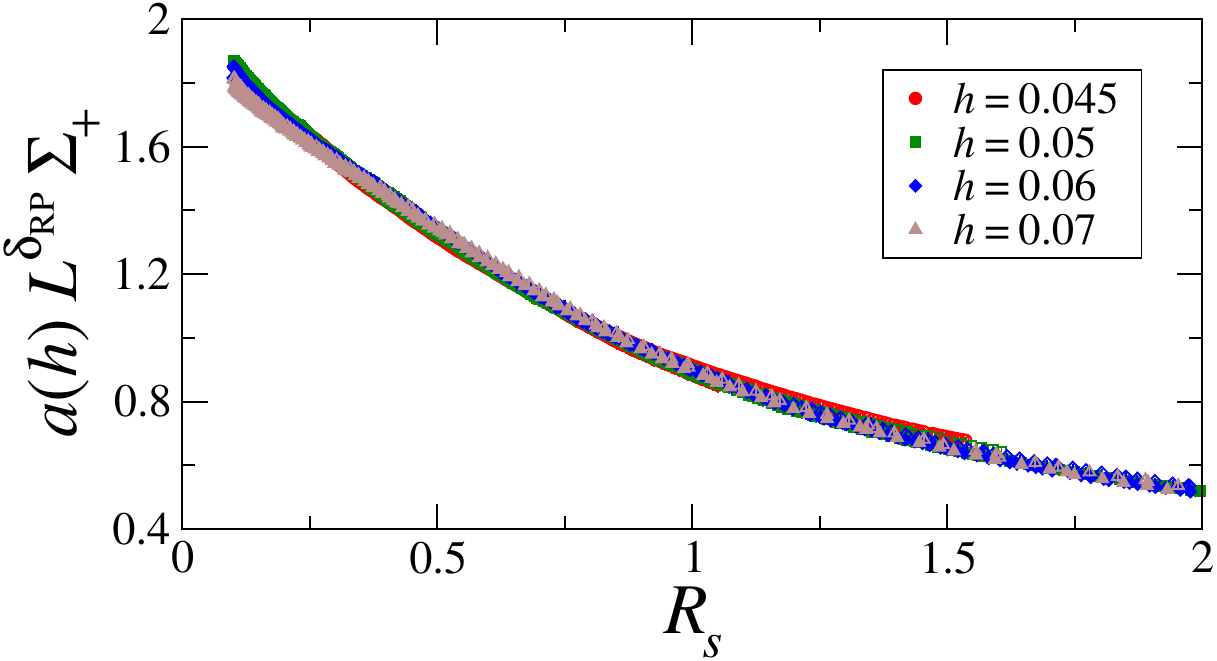}
  \caption{Plot of $a(h) L^{\delta_{\rm RP}} \Sigma_+$ vs $R_s$ for
    several values of $h$, where $a(h)$ is chosen to optimize the data
    collapse. Setting $a = 1$ for $h = 0.045$, we use $a=1.137$,
    $1.096$, $1.037$ for $h = 0.07$, $0.06$, $0.05$, respectively.
    Results are shown for $L = 2000$ (empty symbols) and $L=4000$
    (filled symbols).  The resulting scaling collapse is good, with
    small deviations that can be explained by residual scaling
    corrections.}
\label{cluster2Dglobalscaling}
\end{figure}
%%%%%%%%%%%%%%%%%%%%%%%%%%%%%%%%%%%%%%%%%%%%%%%%%%%%%%%%%%%%%%%%%%%%%%%

The fractal-dimension exponent $\delta \equiv \delta_+=\delta_-$ can
be obtained by matching the data for $\Sigma_\pm(t)$ [see
  Eq.~\eqref{drddef}] with the asymptotic FSS behavior in
Eq.~\eqref{Dpiu-scaling}.  Since ${\cal F}_R(X)$ is monotonically
decreasing, one can invert Eq.~\eqref{Ansatz-FSS} and express $X$ as a
function of $R_s$, obtaining
\begin{equation}
  \Sigma_\pm(t,L) \approx L^{-\delta} \widetilde{\cal F}_\pm(R_s),
  \label{Dpiu-scaling2}
\end{equation}
in which only the exponent $\delta$ appears.  Standard fits to
Eqs.~\eqref{Dpiu-scaling} and~\eqref{Dpiu-scaling2} yield the
estimates reported in Table~\ref{table-td-2D}, which vary very little
around $\delta\approx 0.10$.  We note that these estimates do not
account for possible systematic effects due to scaling corrections,
which become more pronounced as $h$ decreases and may induce
deviations larger than the quoted errors. Remarkably, the numerical
estimates of $\delta$ are very close to the RP value $\delta_{\rm RP}
= 5/48 \approx 0.104$~\cite{Kesten-book,SA-book}, indeed differences
are less than two error bars.  This suggests that the exponent
$\delta$ actually coincides with $\delta_{\rm RP}$.  Moreover, the
geometric properties of the critical clusters apparently exhibit an
$h$-independent universal behavior. In particular, the data provide
evidence of the universality of the FSS function
\begin{equation}
  \widetilde{\cal F}_\pm(R_s) \approx L^{\delta_{\rm RP}} \Sigma_\pm(t,L)
  \label{wtfpmrs}
\end{equation}
in the large $L$ limit (up to a multiplicative normalization factor).
As shown in Fig.~\ref{cluster2Dglobalscaling}, the data for
$L^{\delta_{\rm RP}} \Sigma_\pm(t,L)$ collapse after a trivial
multiplicative rescaling, providing evidence that the scaling
properties are $h$-independent and consistent with those of standard
RP critical clusters, even though the approach to the transition (most
notably the exponent $w$) retains a nontrivial dependence on $h$.

%%%%%%%%%%%%%%%%%%%%%%%%%%%%%%%%%%%%%%%%%%%%%%%%%%%%%%%%%%%%%%%%%%%%%%%
\begin{figure}[tbp]
  \includegraphics[width=0.9\columnwidth]{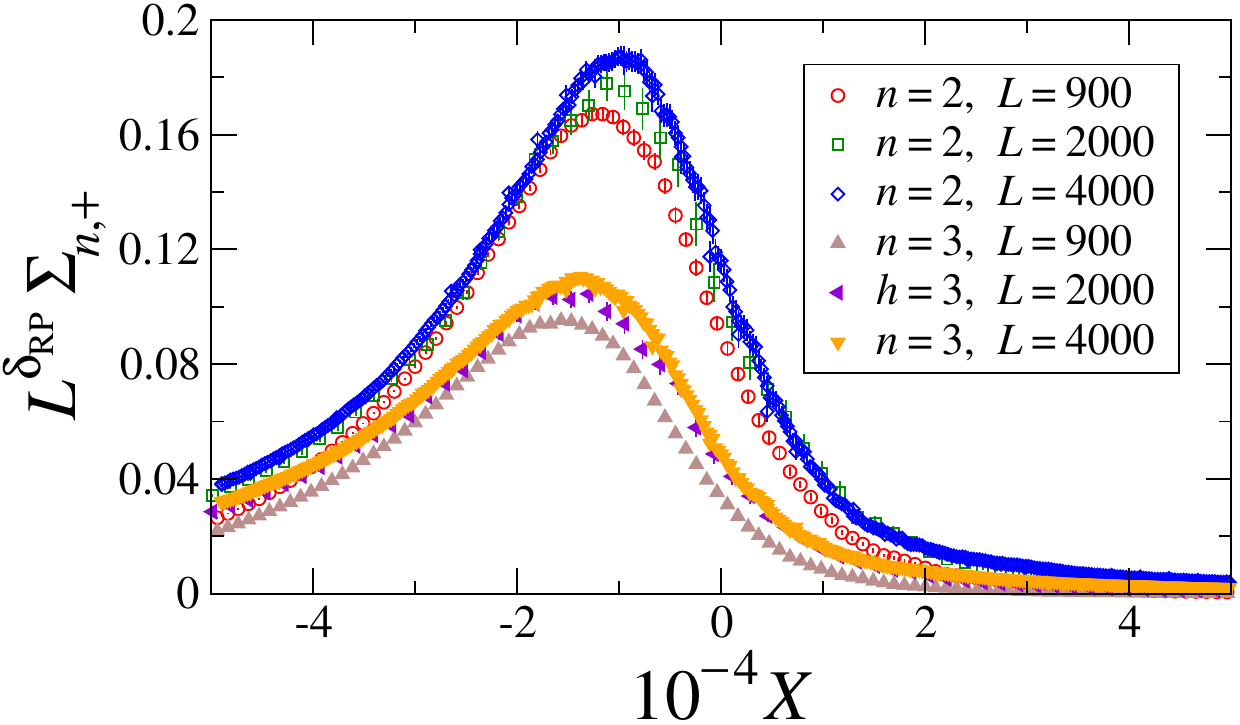}
  \caption{Rescaled cluster sizes $L^{\delta_{\rm RP}} \Sigma_{n,+}$ vs $X
    = (t-t_c) L^{w}$ for $h = 0.07$, with $t_c = 1174$ and $w = 0.68$,
    $\delta_{\rm RP} = 5/48$.  We report results for the second ($n=2$,
    empty symbols) and third ($n=3$, filled symbols) largest clusters.}
  \label{Sigma_23_h0p07}
\end{figure}
%%%%%%%%%%%%%%%%%%%%%%%%%%%%%%%%%%%%%%%%%%%%%%%%%%%%%%%%%%%%%%%%%%%%%%%

Finally, we have also considered the first few $n$th largest
clusters, finding that in all cases their fractal dimension is
compatible with $d_{\rm RP}$ (see Fig.~\ref{Sigma_23_h0p07} for
results for $n=2$ and 3 at $h = 0.07$). Thus, in the infinite-volume
limit at $t = t_c$, the system is characterized be an infinite number
of positive-spin and negative-spin domains, all having the same
fractal dimension $d_{\rm RP}$.

\section{Post-quench magnetization in the thermodynamic limit}
\label{infmag}

\subsection{The infinite-volume magnetization}

%%%%%%%%%%%%%%%%%%%%%%%%%%%%%%%%%%%%%%%%%%%%%%%%%%%%%%%%%%%%%%%%%%%%%%%
\begin{figure}[tbp]
  \includegraphics[width=0.9\columnwidth]{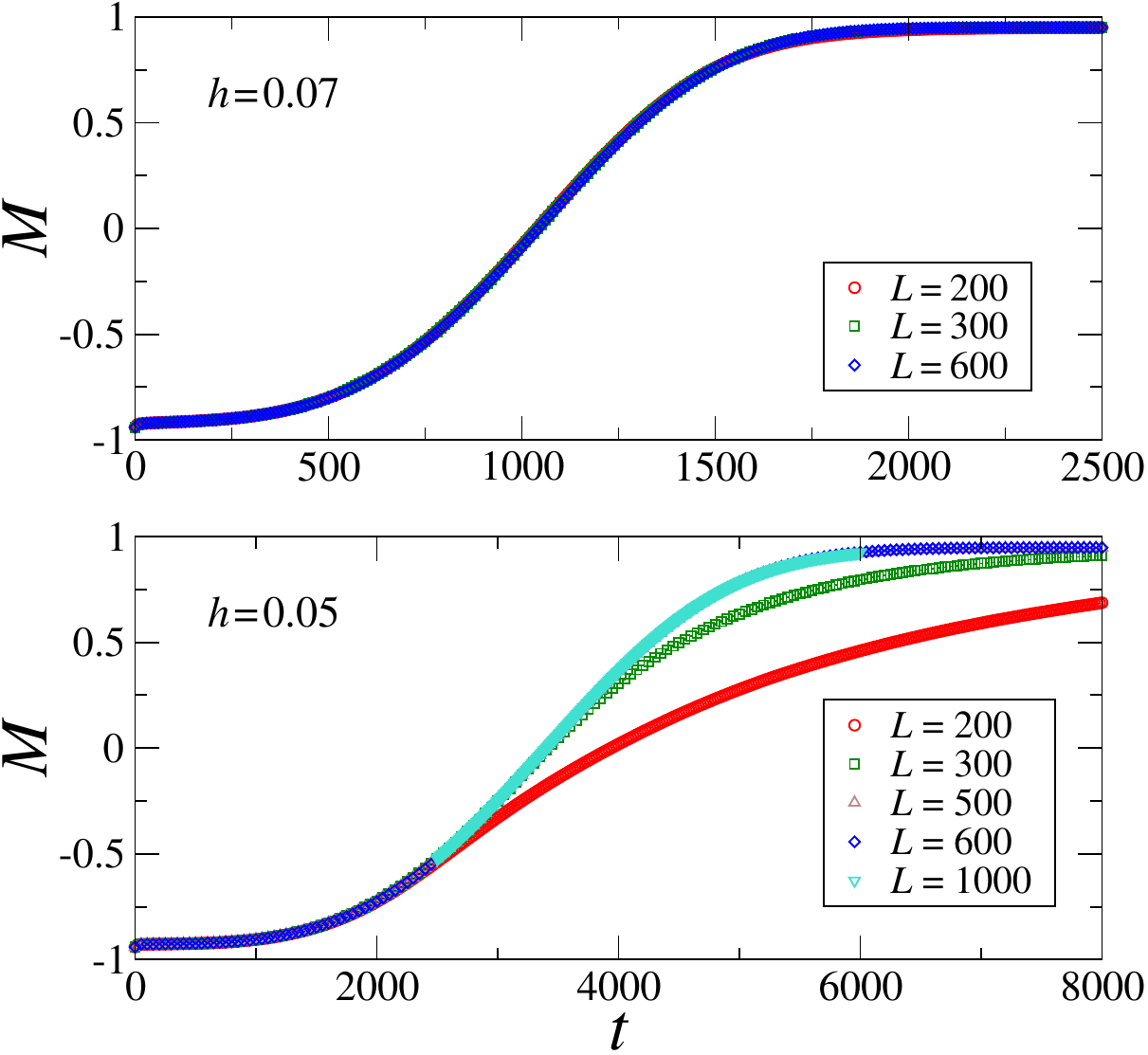}
  \caption{Time evolution of the magnetization $M(t)$ for $h=0.07$
    (top) and $h = 0.05$ (bottom) and several sizes $L$.  Statistical
    errors are hardly visible (for $L=600$, errors are at most
    $0.003$ and $0.004$ for $h=0.07$ and $h=0.05$, respectively).
    Data clearly converge to a limiting curve as $L$ increases, which
    provides a good approximation of $M_\infty(t)$. }
\label{rawlts-suppl}
\end{figure}
%%%%%%%%%%%%%%%%%%%%%%%%%%%%%%%%%%%%%%%%%%%%%%%%%%%%%%%%%%%%%%%%%%%%%%%

%%%%%%%%%%%%%%%%%%%%%%%%%%%%%%%%%%%%%%%%%%%%%%%%%%%%%%%%%%%%%%%%%%%%%%%
\begin{figure}[tbp]
  \includegraphics[width=0.9\columnwidth]{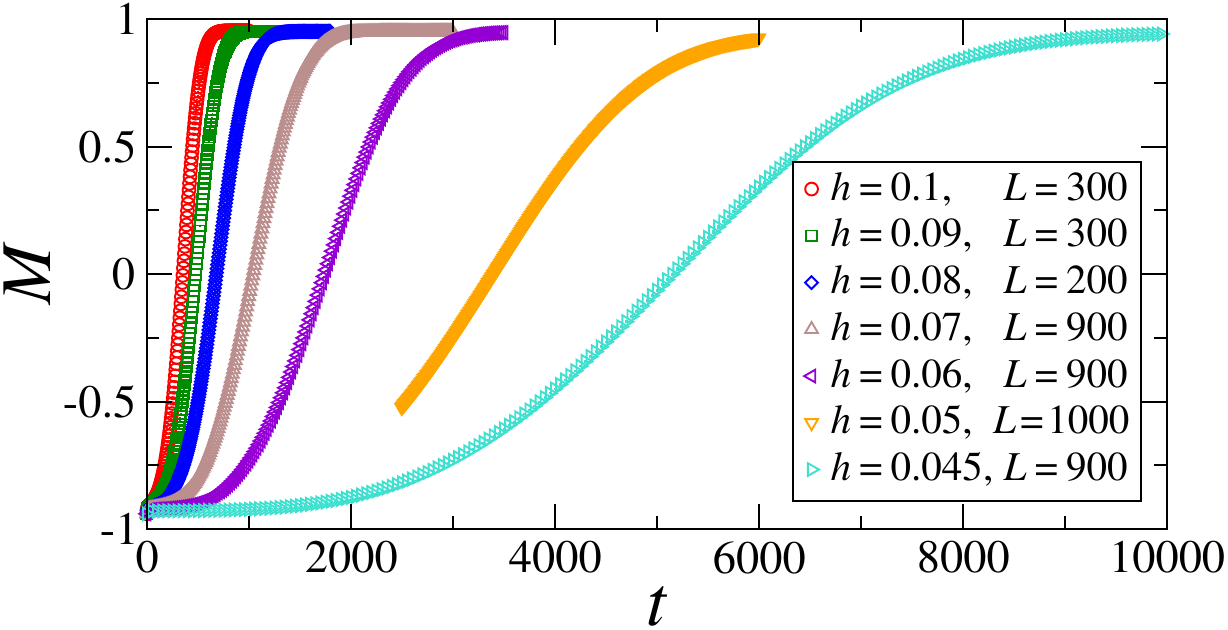}
  \caption{Time dependence of $M(t)$ for several values of $h$, and
    sufficiently large $L$ to provide their infinite-volume limit
    $M_\infty(t)$ (statistical errors are hardly visible).}
\label{magn-allh-suppl}
\end{figure}
%%%%%%%%%%%%%%%%%%%%%%%%%%%%%%%%%%%%%%%%%%%%%%%%%%%%%%%%%%%%%%%%%%%%%%%

We now analyze the scaling behavior of the infinite-volume
magnetization $M_\infty(t)$ at fixed $h$.  To determine it for a given
value of $h$, we increase the size $L$ until the average curves of
$M(t)$ become approximately $L$-independent (in practice, equal within
the statistical errors).  The curves corresponding to the largest $L$
provide the infinite-size values $M_\infty(t)$, for the given value of
$h$.  As an example, in Fig.~\ref{rawlts-suppl} (top) we show the
magnetization data for two values of $h$ and different lattice
sizes. For $h = 0.07$, simulations with $L=200$ already provide the
infinite-volume magnetization curves. As $h$ decreases, finite-size
effects become larger.  For example, for $h=0.05$ [see
  Fig.~\ref{rawlts-suppl} (bottom)], the large-$L$ limit (within
errors) is obtained for $L\gtrsim 600$ ($L\gtrsim 900$, for
$h=0.045$). This behavior represents the major limitation to the
possible values of $h$ that we can probe in the small-$h$ regime.

The infinite-volume results for different values of $h$ in the
interval $[0.045,0.10]$ are reported in Fig.~\ref{magn-allh-suppl}.
For $h = 0.10$, the transition to the positively magnetized phase
occurs rapidly and indeed, for larger values of $h$, it is difficult
to identify an out-of-equilibrium regime. As $h$ decreases, the system
remains in the negatively magnetized state over progressively longer
time intervals.

\begin{figure}[tbp]
  \includegraphics[width=0.9\columnwidth]{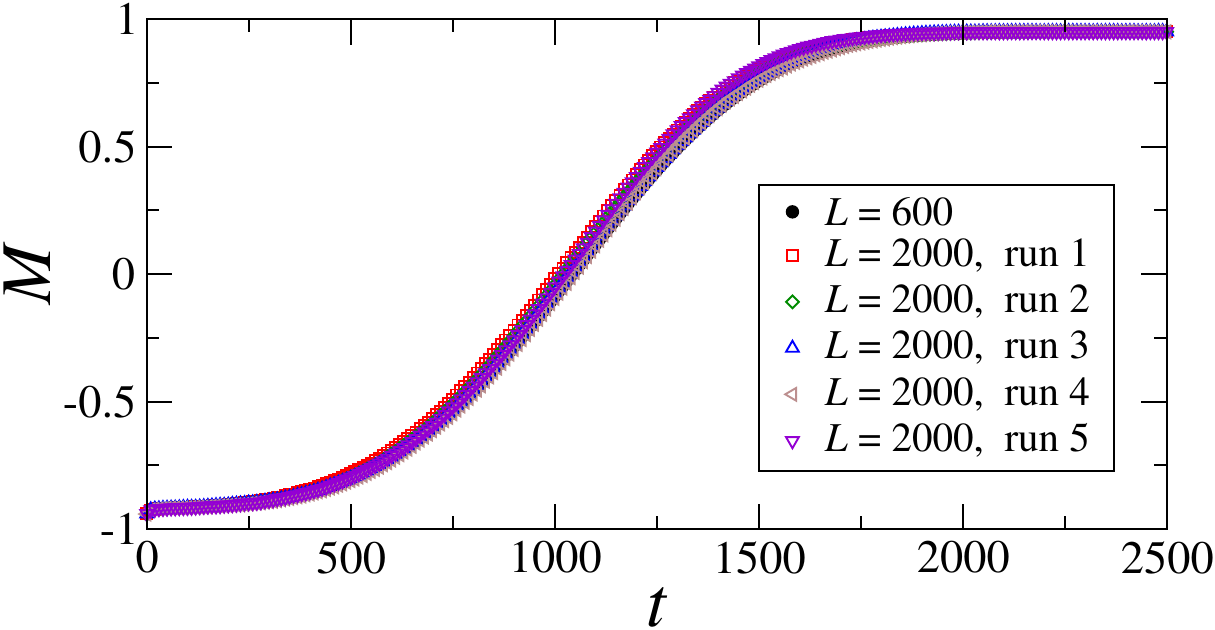}
  \caption{Comparison of the average magnetization $M(t)$ for $L=600$
    with the instantaneous magnetization for five different runs on a
    large lattice with $L=2000$, for $h=0.07$.}
  \label{magn-singoli-suppl}
\end{figure}

We should stress that the infinite-volume magnetization curves
do not only provide information on the average behavior of the system
as a function of $t$, but also represent the typical evolution of any
sufficiently large system.  Indeed, self-averaging holds, so that
fluctuations in the dynamical evolution vanish for $L\to \infty$.
This is illustrated in Fig.~\ref{magn-singoli-suppl}, where we compare
the evolution of the instantaneous magnetization for large systems
(here $L=2000$) with the average behavior computed on smaller
lattices.

%%%%%%%%%%%%%%%%%%%%%%%%%%%%%%%%%%%%%%%%%%%%%%%%%%%%%%%%%%%%%%%%%%%%%%%
%%%%%%%%%%%%%%%%%%%%%%%%%%%%%%%%%%%%%%%%%%%%%%%%%%%%%%%%%%%%%%%%%%%%%%%

\subsection{Scaling behavior of the magnetization} \label{sec2.B1}

A general scaling form for the time-dependent magnetization in Ising
systems driven across the magnetic FOT line was conjectured in
Ref.~\cite{PV-26}. We briefly review the general argument, as it plays
a crucial role in the analysis of the magnetization under a magnetic
quench.  We assume that, for the infinite-volume dynamics across the
FOT line, the relevant scaling variable is the magnetic energy of
droplets of positive spins nucleated during the dynamical evolution,
which are eventually responsible for the transition from the
metastable negatively magnetized state to the positively magnetized
one.  Under this hypothesis, the relevant scaling variable is
expected to be
\begin{equation}
  \sigma \sim R(t)^d h,
  \label{sigmafr}
\end{equation}
where $R(t)$ denotes the typical droplet size at time $t$, and $d$ is
the spatial dimension. Assuming droplets to have smooth boundaries,
the time needed to form a droplet of size $R$ should scale as $\exp(c
R^{d-1})$ in $d$ dimensions, implying
\begin{equation}
  R(t) \sim (\ln t)^{1/(d-1)},
  \label{rtlnt}
\end{equation}
i.e., $R(t) \sim \ln t$ in two dimensions.  We thus predict that the
relevant scaling variable is
\begin{equation}
  \sigma = (\ln t)^2 h .
  \label{sdef}
\end{equation}
It is thus natural to conjecture that the 
infinite-volume magnetization $M_\infty(t,h)$ scales as 
\begin{equation}
  M_\infty(t,h) \approx {\cal M}_\infty(\sigma).
 \label{Mscaling}
\end{equation}

\begin{figure}[tbp]
  \includegraphics[width=0.9\columnwidth]{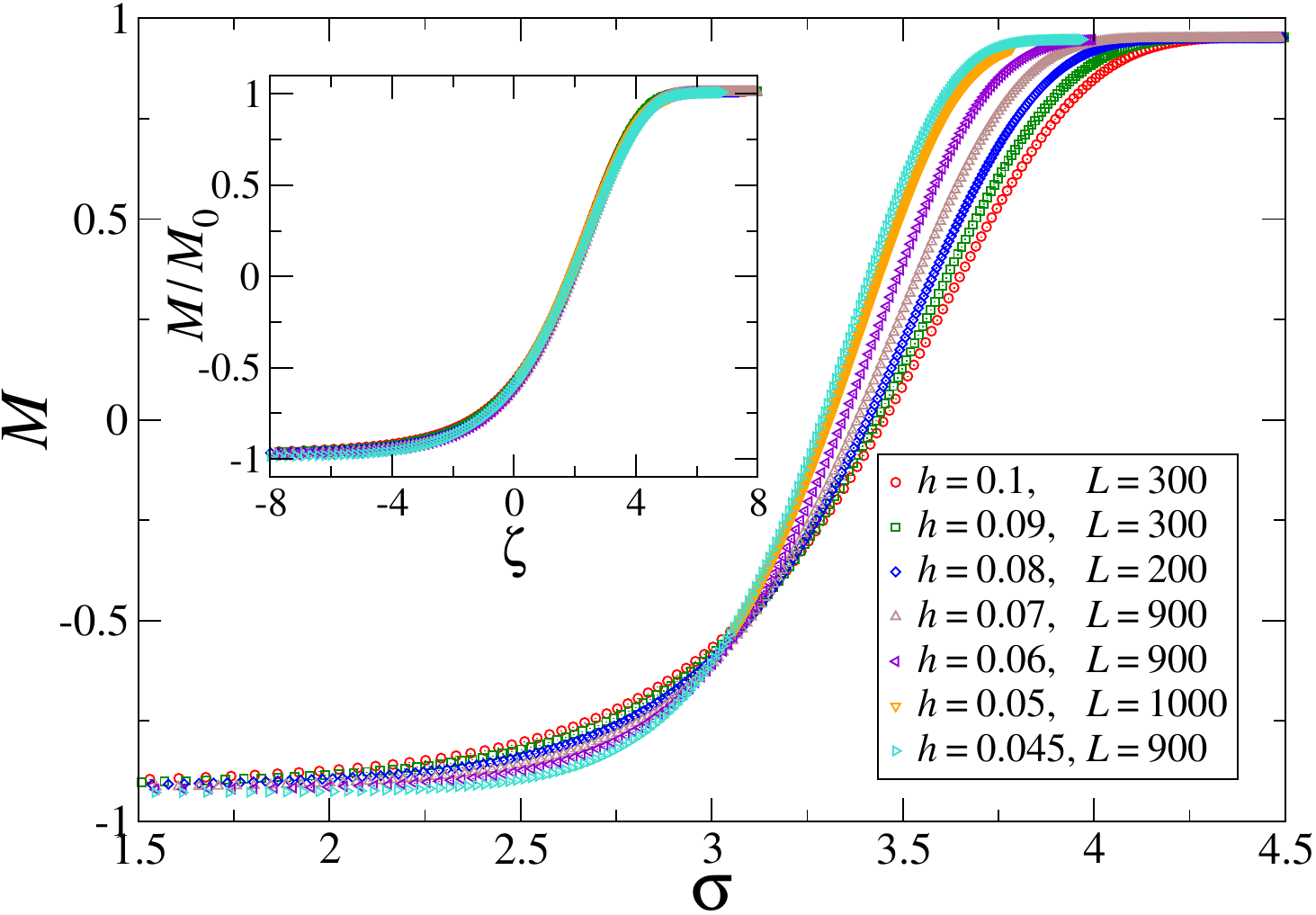}
  \caption{The magnetization $M(t)$ plotted vs $\sigma = (\ln t)^2h$.
    The size $L$ is large enough that the data can be regarded in
    their thermodynamic limit. The inset shows the ratio $M(t)/M_0$ vs
    $\zeta=(\sigma-\sigma_*) h^{-\theta}$ with $\theta=0.64$ and
    $\sigma_*= 3.03$, where $M_0\approx 0.940259$ is the spontaneous
    magnetization for $\beta = 1.2\beta_c$ [cf.~Eq.~\eqref{m02d}].}
\label{resc}
\end{figure}

This is confirmed by the numerical data.  As shown in Fig.~\ref{resc},
the magnetization $M_\infty(t,h)$ for different values of $h$ get
closer and closer as a function of $\sigma$, as $h$
decreases. Moreover, the curves intersect at $\sigma=\sigma_*\approx
3.0$.  Around $\sigma_*$, we find an additional scaling behavior
\begin{equation}
  M_\infty(t,h) \approx \widetilde{\cal M}_\infty(\zeta), \quad
  \zeta = (\sigma-\sigma_*) \, h^{-\theta},
  \label{estar}
\end{equation}
where $\theta>0$ (see the inset of Fig.~\ref{resc}).  Fits to the data
with Eq.~\eqref{estar} yield
\begin{equation}
  \sigma_*=3.03(3),\qquad \theta = 0.64(4).
  \label{sigstarthe}
\end{equation}  
The scaling form~\eqref{estar} implies that the time dependence of the
magnetization $M_\infty$, in the limit $h\to 0$, develops a
discontinuity at $\sigma_*$ as a function of $\sigma$, i.e.,
\begin{equation}
  \lim_{\sigma\to\sigma_*^\pm} M_\infty(\sigma) \approx \pm M_0,
  \label{discmt}
\end{equation}
where $M_0$ is the spontaneous magnetization~\eqref{m02d}.  The
approach to these limiting values is controlled by corrections
decreasing as $h^{\theta}$.

%%%%%%%%%%%%%%%%%%%%%%%%%%%%%%%%%%%%%%%%%%%%%%%%%%%%%%%%%%%%%%%%%%%%%%%
%%%%%%%%%%%%%%%%%%%%%%%%%%%%%%%%%%%%%%%%%%%%%%%%%%%%%%%%%%%%%%%%%%%%%%%

The time scale corresponding to the crossing value $\sigma_*$,
\begin{equation}
  t_*(h) = e^{\sqrt{\sigma_*/h}},
  \label{tstarh}
\end{equation}
turns out to be related to the percolation time $t_c(h)$. In
Table~\ref{table-td-2D} we report the related quantities
\begin{equation}
  \sigma_c \equiv  (\ln t_c)^2 h,\qquad \zeta_c \equiv (\sigma_c - \sigma_*)
  h^{-\theta},
  \label{sigczec}
\end{equation}
showing that $\zeta_c$ is approximately constant within the errors.
This implies that
\begin{equation}
  t_*(h) \approx t_c(h) (1 - b h^\theta), \qquad b =
  {\zeta_c\over 2 \sigma_*} = 0.45(2).
  \label{tstarh2}
\end{equation}
Therefore, for $h\to 0$, also
$t_c$ diverges as
\begin{equation}
  t_c(h) \approx t_*(h) = e^{\sqrt{\sigma_*/h}}.
  \label{tcsca}
\end{equation}
The exponent $\theta$ controls the approach to $h=0$. It is thus
natural to conjecture that the $h$-dependence of $w$ is also
controlled by $\theta$, implying $w\propto h^\theta$. Data are
consistent with this scenario: fits of $w(h)$ to the form $a h^\alpha$
yield $\alpha = 0.60(2)$ (including all values of $h$), and $\alpha =
0.72(8)$ (discarding the result for $h = 0.07$).

Note that $t_*$ is significantly shorter than the time
scale obtained by requiring isolated droplets to be energetically
stable. Indeed, the energy of a positive-spin droplet surrounded by
negative spins is
\begin{equation}
  E = - a h R^d + b R^{d-1}, \qquad a,b>0.
  \label{droene}
\end{equation}
It is immediate to verify that the energy decreases with increasing
$R$, only for $R$ larger than a threshold $R_c$ that scales as
$1/h$. Droplets of size $R_c$ require a time $t_D$ of order
\begin{equation}
  t_D \sim  e^{c R_c} \sim e^{\bar{c}/h}
  \label{expcrc}
\end{equation}
to be generated~\cite{RTMS-94}.  For $h\to 0$, this time scale is
significantly larger than the one defined in Eq.~\eqref{tcsca},
indicating that in 2D Ising systems the transition occurs when typical
isolated droplets are too small to be energetically stable. Thus, the
dominant mechanism responsible for the transition in 2D Ising systems
is not the independent growth of isolated droplets, but rather their
collective aggregation dynamics: droplets increase their size
predominantly by merging, a dynamical process that is always
energetically favored.  This mechanism allows the system to generate
stable positive-spin domains on time scales much shorter than $t_D
\sim e^{\bar{c}/h}$, and is responsible for the percolation transition
that marks the passage from one magnetic phase to the other.

The relevance of the merging dynamics in the critical region is
particularly evident in Fig.~\ref{clusterfig2D}, which shows
configurations of the system at three different times within the
critical region: we consider $t=t_c-30, \, t_c, \, t_c+30$, with $t_c
= 1174$ (results are for $h=0.07$). Over this short time window, the
cluster boundaries remain essentially unchanged, while the size
$\Sigma_+$ of the largest positive cluster increases significantly,
from 0.37 (at $t=t_c-30$), to 0.50 (at $t=t_c$), and 0.64 (at $t=t_c+30$).
The growth is thus not driven by interface motion, but by the merging
of the largest cluster with smaller positive-spin clusters.

\section{Quench dynamics in systems with fixed boundary conditions}
\label{sec4}

In the previous section we have studied the scaling behavior of the
infinite-volume magnetization determined from finite-size data
computed on lattices with PBC. We now verify that such infinite-volume
results are independent of the boundary conditions, a result which is
not completely obvious since, in the infinite-volume limit the
equilibrium behavior at FOTs does depend on the boundary conditions,
which may stabilize a single phase or force the coexistence of
different phases.  We consider FBC with $s_{\bm x} = 1$ on the lattice
boundary and compare the resulting behavior with that obtained using
PBC. The quenching protocol is the same as that outlined in
Sec.~\ref{quepro}, i.e., it starts from negatively magnetized
configurations with $s_{\bm x}=-1$ at any lattice site within the
boundary, and the post-quench phase change is driven by a positive
magnetic field $h>0$.  To speed up simulations, here we consider the
Metropolis dynamics and determine the evolution of the system for
$h=0.09,\,0.07,\,0.06$.

%%%%%%%%%%%%%%%%%%%%%%%%%%%%%%%%%%%%%%%%%%%%%%%%%%%%%%%%%%%%%%%%%%%%%%%
\begin{figure}[tbp]
  \includegraphics[width=0.9\columnwidth]{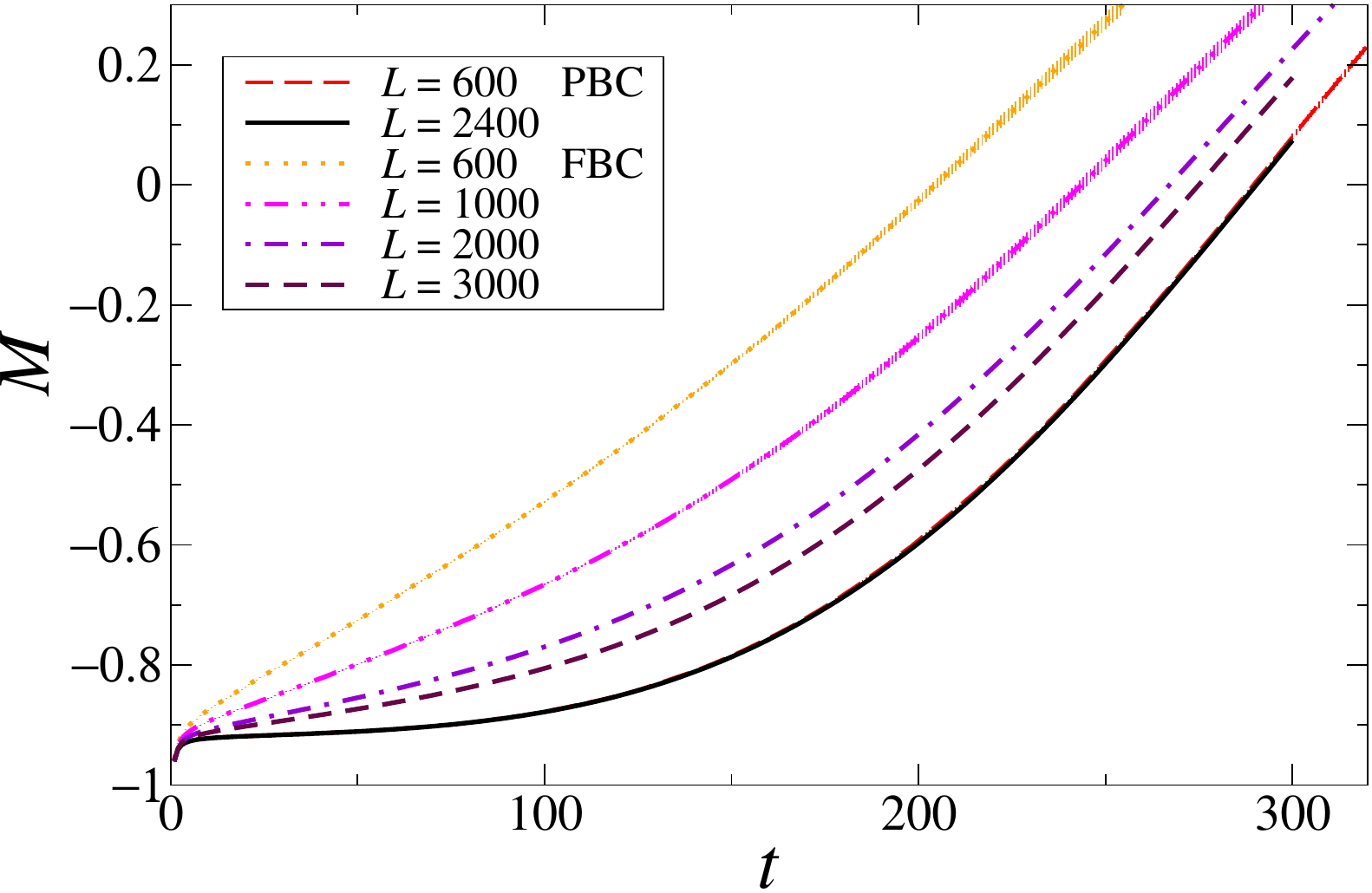}
  \caption{Time dependence of the post-quench magnetization for
    $h=0.07$, for systems with PBC and FBC and various values of $L$.
    Statistical errors are at most of the size of the thickness of the
    lines.}
\label{2Dmt}
\end{figure}
%%%%%%%%%%%%%%%%%%%%%%%%%%%%%%%%%%%%%%%%%%%%%%%%%%%%%%%%%%%%%%%%%%%%%%%

In Fig.~\ref{2Dmt} we show the time-dependent magnetization for
$h=0.07$ and different sizes.  As expected, systems with FBC
anticipate the phase change with respect to those with PBC,
essentially because positively magnetized boundary conditions
favor the passage from the negatively to the positively magnetized
phase. While with PBC data for $L=600$ and $2400$ coincide within
errors (convergence to the infinite-volume limit is likely exponential),
with FBC the magnetization curves show a significant dependence
on $L$. This is not unexpected as, in the latter case, one expects
corrections to decay slower as $1/L$ for global quantities like
the magnetization. Similar results are obtained for
other values of $L$.

%%%%%%%%%%%%%%%%%%%%%%%%%%%%%%%%%%%%%%%%%%%%%%%%%%%%%%%%%%%%%%%%%%%%%%%
\begin{figure}[tbp]
  \includegraphics[width=0.9\columnwidth]{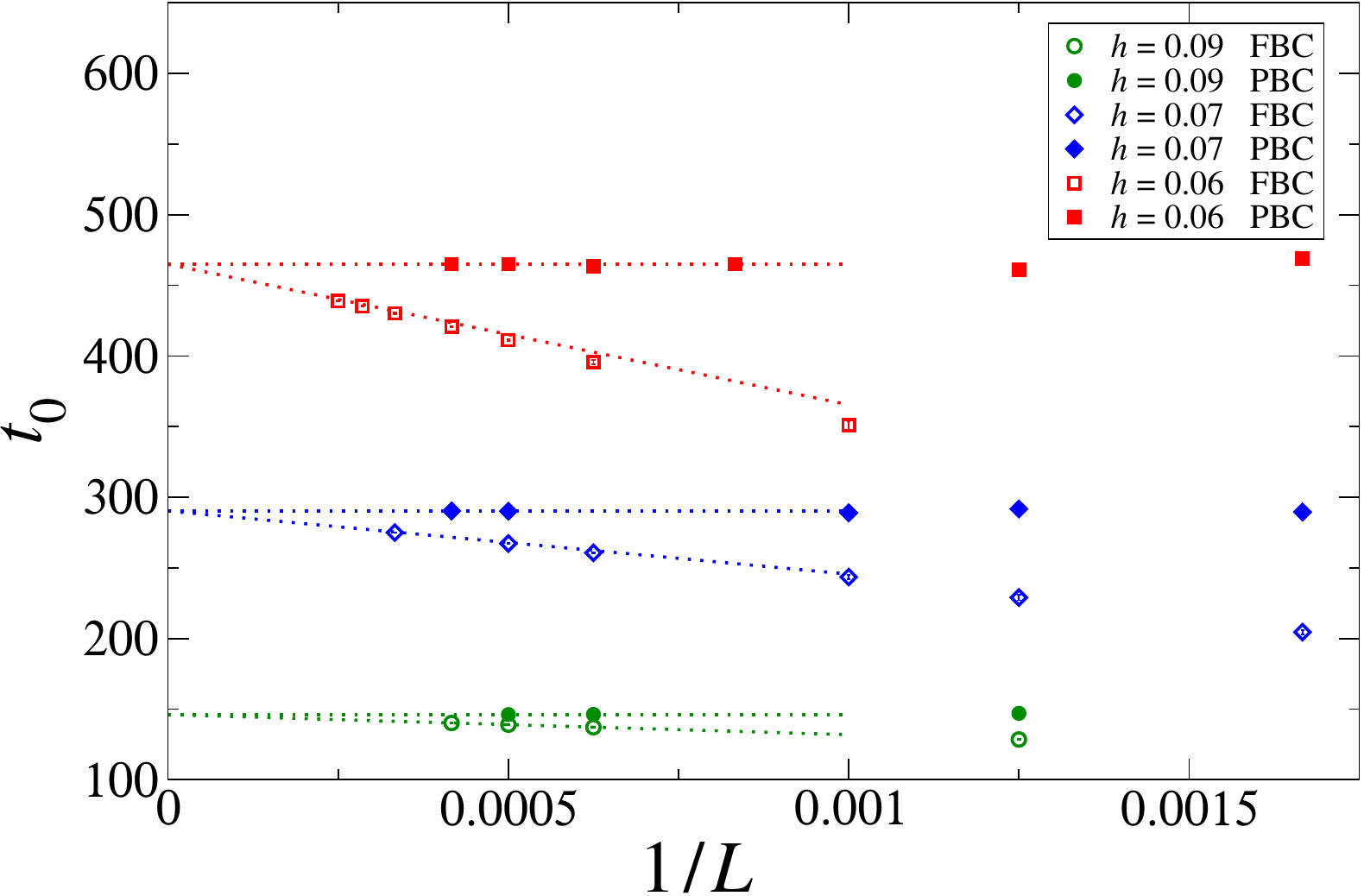}
  \caption{Estimates of the time $t_0$ at which the post-quench
    magnetization vanishes. Results are for FBC (empty)
    and for PBC (filled).  Dotted lines are drawn to guide the eye.}
\label{2Dt0}
\end{figure}
%%%%%%%%%%%%%%%%%%%%%%%%%%%%%%%%%%%%%%%%%%%%%%%%%%%%%%%%%%%%%%%%%%%%%%%

%%%%%%%%%%%%%%%%%%%%%%%%%%%%%%%%%%%%%%%%%%%%%%%%%%%%%%%%%%%%%%%%%%%%%%%
\begin{figure}[tbp]
  \includegraphics[width=0.9\columnwidth]{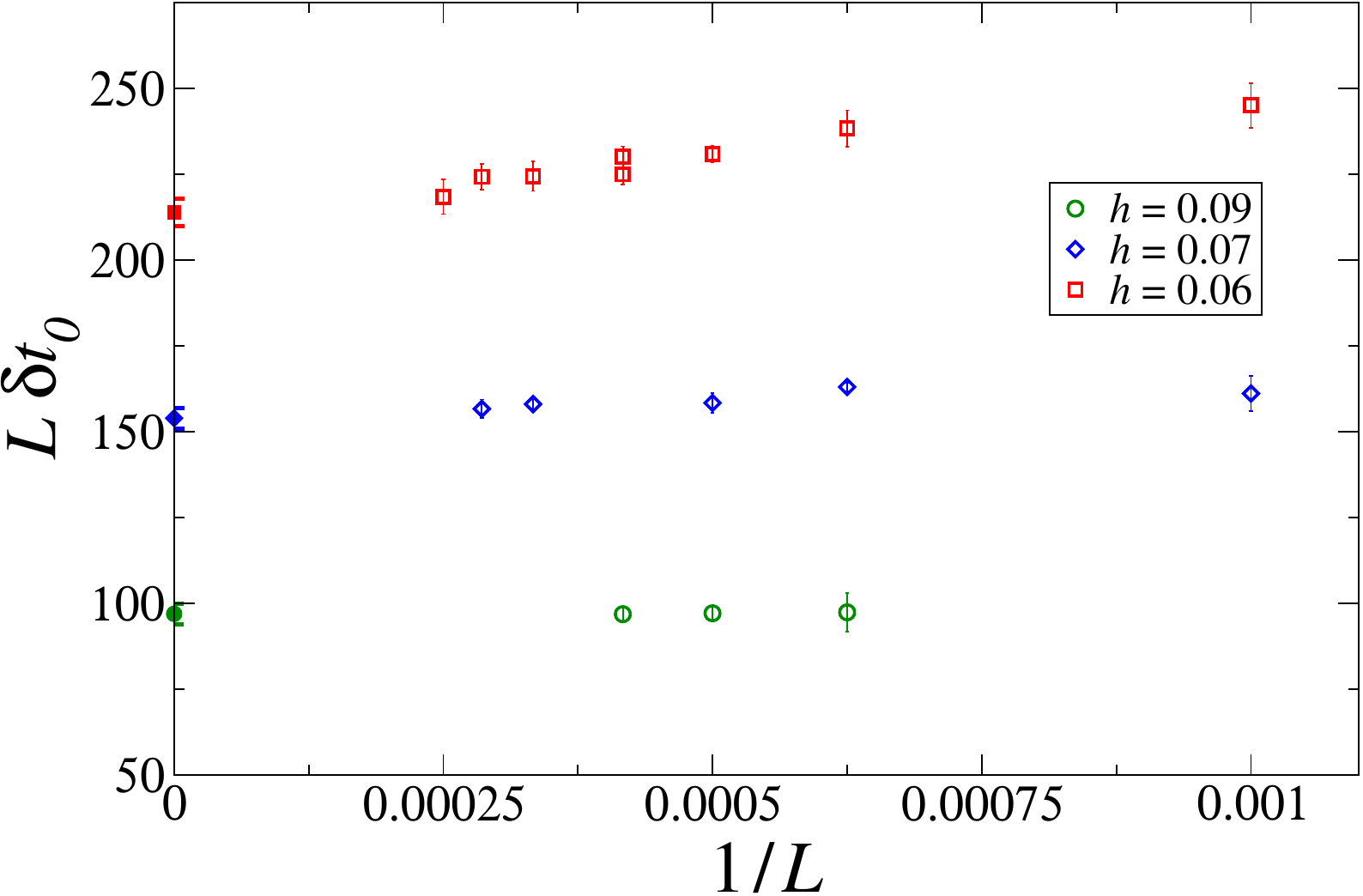}
  \caption{Plots of $\delta t_0 L$, providing an approximation of
    the $1/L$ coefficient $c(h)$ defined in Eq.~\eqref{t0fbc}, and of
    their extrapolations (results on the left vertical axis) using the
    Ansatz in Eq.~\eqref{fitlt0}.}
\label{2Dt0c}
\end{figure}
%%%%%%%%%%%%%%%%%%%%%%%%%%%%%%%%%%%%%%%%%%%%%%%%%%%%%%%%%%%%%%%%%%%%%%%

Apparently we observe that the FBC magnetization curves get closer
to the PBC result as $L$ increases. To make a more quantitative
comparison and verify that the results coincide for $L\to \infty$, we
determine the time $t_0$ at which the magnetization vanishes in the
post-quench evolution, i.e., when $M(t_0) = 0$.  Estimates of $t_0$
for different values of $h$ and $L$ are shown in Fig.~\ref{2Dt0}.
They clearly have a finite limit for $L\to\infty$, but with
significant $1/L$ corrections, as expected. Most importantly,
the limiting value for $L\to \infty$ is perfectly consistent with
the PBC infinite-volume estimate.  More precisely, we find that
\begin{equation}
  \delta t_0 = {t_{0,\infty}-t_{0,{\rm fbc}}(L) \over t_{0,\infty}}
  \approx {c(h)\over L}.
  \label{t0fbc}
\end{equation}
The value $t_{0,\infty}$ can be straightforwardly estimated from the
large-size PBC data, obtaining
$t_{0,\infty} = 146.1(1)$, $290.3(1)$, $465.0(3)$ for
$h = 0.09$, $0.074$, $0.06$, respectively (we interpolated data
for integer $t$).

The $1/L$ approach of FBC data is clearly
demonstrated by the plots of $L\,\delta t_0$ reported in
Fig.~\ref{2Dt0c}, which provide approximations of the $1/L$
coefficient $c(h)$ defined in Eq.~\eqref{t0fbc}.  Assuming
next-to-leading size corrections of order $L^{-2}$, we can fit the
data of $L\,\delta t_0$ to
\begin{equation}
  L\,\delta t_0 =  c(h) + {c_1(h)\over L},
  \label{fitlt0}
\end{equation}
obtaining $c(h=0.09) = 97(3)$,
$c(h=0.07) = 154(3)$, and $c(h=0.06) = 214(4)$. The coefficient $c(h)$
is rapidly increasing with decreasing $h$. In particular, the results
are numerically consistent with the power law
\begin{equation}
  c(h) = \hat{c} \,h^{-2},\qquad \hat{c} = 0.77(2).
  \label{chbeh}
\end{equation}
This indicates that the convergence is nonuniform for $h\to 0^+$,
which is not unexpected since the infinite-size limit at a FOT
crucially depends on the boundary conditions.

%%%%%%%%%%%%%%%%%%%%%%%%%%%%%%%%%%%%%%%%%%%%%%%%%%%%%%%%%%%%%%%%%%%%%%%
\begin{figure}[tbp]
  \centering
  \includegraphics[width=0.98\columnwidth]{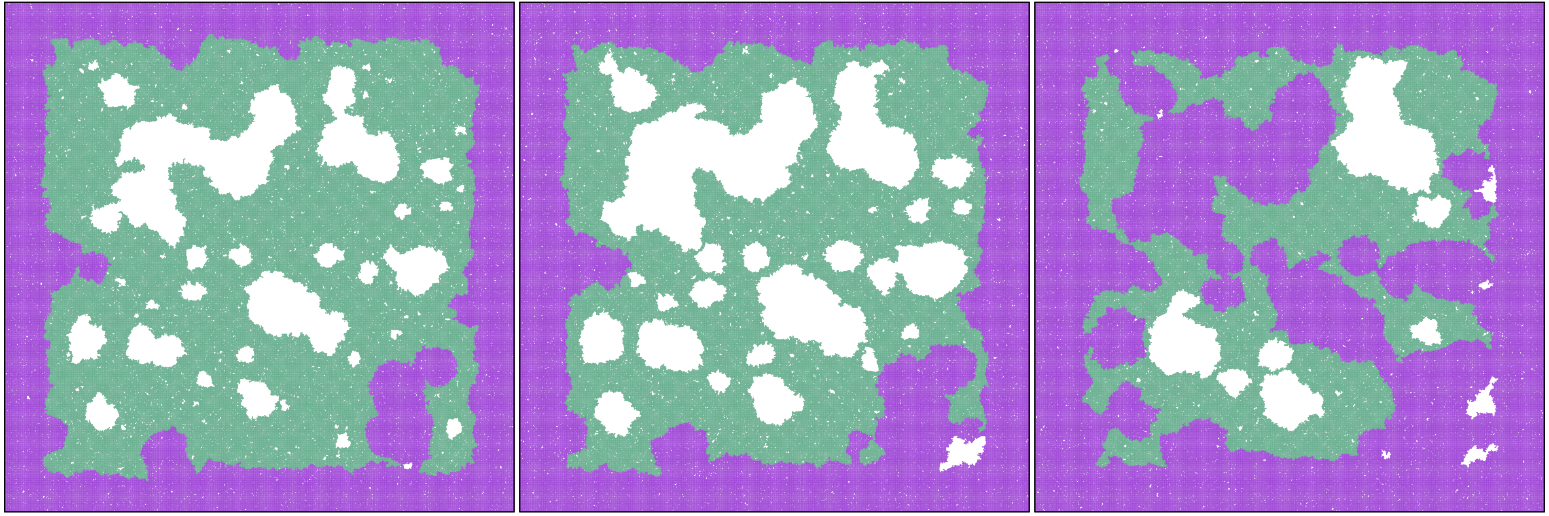}
  \vskip 2mm
  \includegraphics[width=0.98\columnwidth]{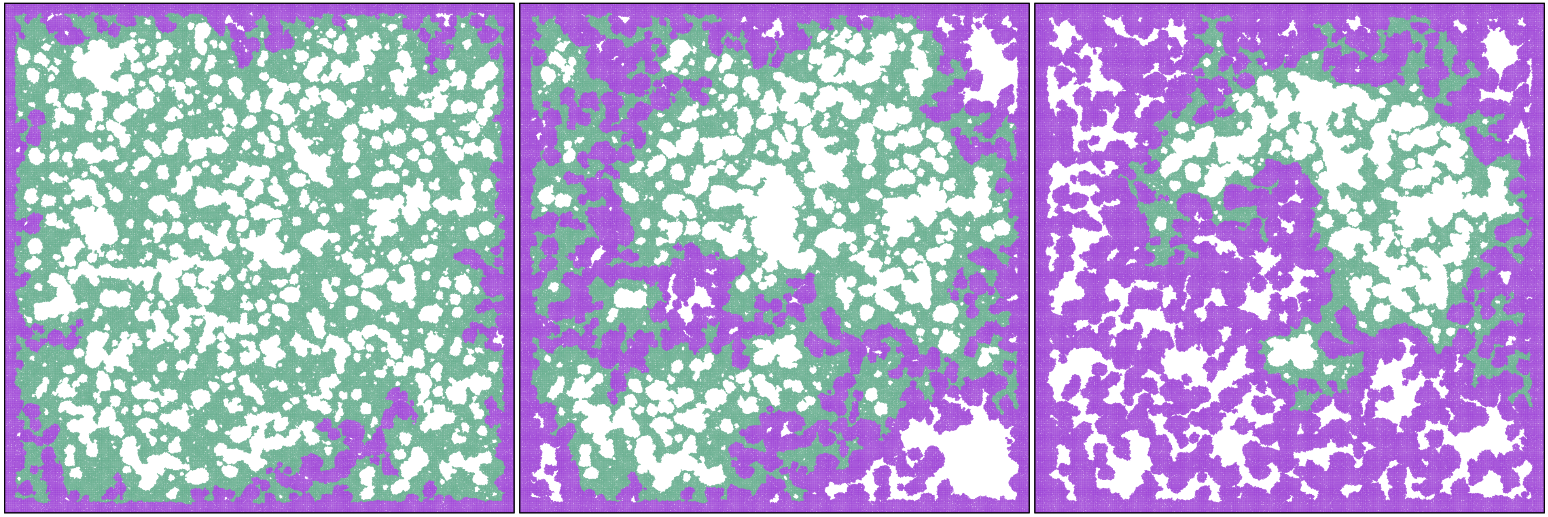}
  \caption{Snapshots of the configurations for $h = 0.07$ at different
    times in the transition region: $M(t) = 0$ (left),
    $0.2$ (middle), and $0.4$ (right).
    The top row is for a system of size $L=1000$, while the bottom row
    is for $L=4000$.  Violet and green sites correspond to the largest
    clusters of positive and negative spins, respectively, while white
    regions denote smaller (both positive and negative) clusters.}
\label{clusterfigFBC}
\end{figure}
%%%%%%%%%%%%%%%%%%%%%%%%%%%%%%%%%%%%%%%%%%%%%%%%%%%%%%%%%%%%%%%%%%%%%%%

The independence of the behavior of the magnetization indicates that
the physical mechanism at the basis of the phase change is the same
for PBC and FBC.  This is quite evident from the configuration
snapshots reported in Fig.~\ref{clusterfigFBC}.  Also for FBC the
transition from one phase to the other is due to the rapid merging of the 
positive clusters that grow inside the lattice far from the boundaries,
and it is not due to the slow growth of the stable cluster connected with
the boundary. With respect to the PBC case, the only difference is the
presence of a positively magnetized region along the boundary of width
$w(t)$, which is essentially independent of $L$. This region, which gives 
a contribution of order $1/L$ to the magnetization, is
responsible for the corrections of order $1/L$ that we found
numerically.

\section{Slow dynamics across the first-order transition line}
\label{slowdyn}

\subsection{The protocol}

We now consider a different protocol in which the magnetic field is
increased slowly across the FOT at fixed $\beta > \beta_c$.
Namely, we vary $h$ linearly in time as
\begin{equation}
  h(t) = t/t_s,
  \label{kzht}
\end{equation}  
where $t_s$ denotes the time scale. In the limit of large $t_s$ the
variations of $h$ become very slow.  Again, the time variable $t$
is increased by one after each complete lattice sweep.
The dynamics starts from thermalized configurations at time $t = -0.2
\,t_s$, i.e., in the negatively magnetized phase.  Again, the protocol
is done at fixed temperature: we consider $\beta = 1.2\,\beta_c$ and
some different large values of $t_s$, performing MC simulations on
lattices up to $L=4000$. Averages are performed over different
trajectories ranging from $N_{\rm traj} = 20000$ for the smallest
lattice size, to $N_{\rm traj} = 500$ for $L=4000$.

Protocols entailing a slow crossing of the phase transition, with a
linear time dependence of the Hamiltonian parameters, have been
already considered in several studies of the out-of-equilibrium
dynamics arising when crossing phase transitions, including the limit
of extremely slow variations, $t_s\to \infty$.  In particular, at
continuous transitions they give rise to the so-called Kibble-Zurek
(KZ) dynamics and productions of
defects~\cite{Kibble-80,Zurek-96,PSSV-11,CEGS-12}. Analogous KZ
protocols have been also studied at classical and quantum FOTs (see,
e.g., Refs.~\cite{RV-21,PV-24} and references therein). At FOTs more
complex behaviors emerge, in particular qualitatively different
mechanisms work in a finite volume and in the thermodynamic limit,
giving rise to unrelated scaling behaviors in the two
cases~\cite{PV-17,TV-22,PRV-25,PV-26,PRV-26}.

\subsection{Percolation transition for fixed values of $t_s$}

%%%%%%%%%%%%%%%%%%%%%%%%%%%%%%%%%%%%%%%%%%%%%%%%%%%%%%%%%%%%%%%%%%%%%%%
\begin{figure}[t]
  \includegraphics[width=0.9\columnwidth]{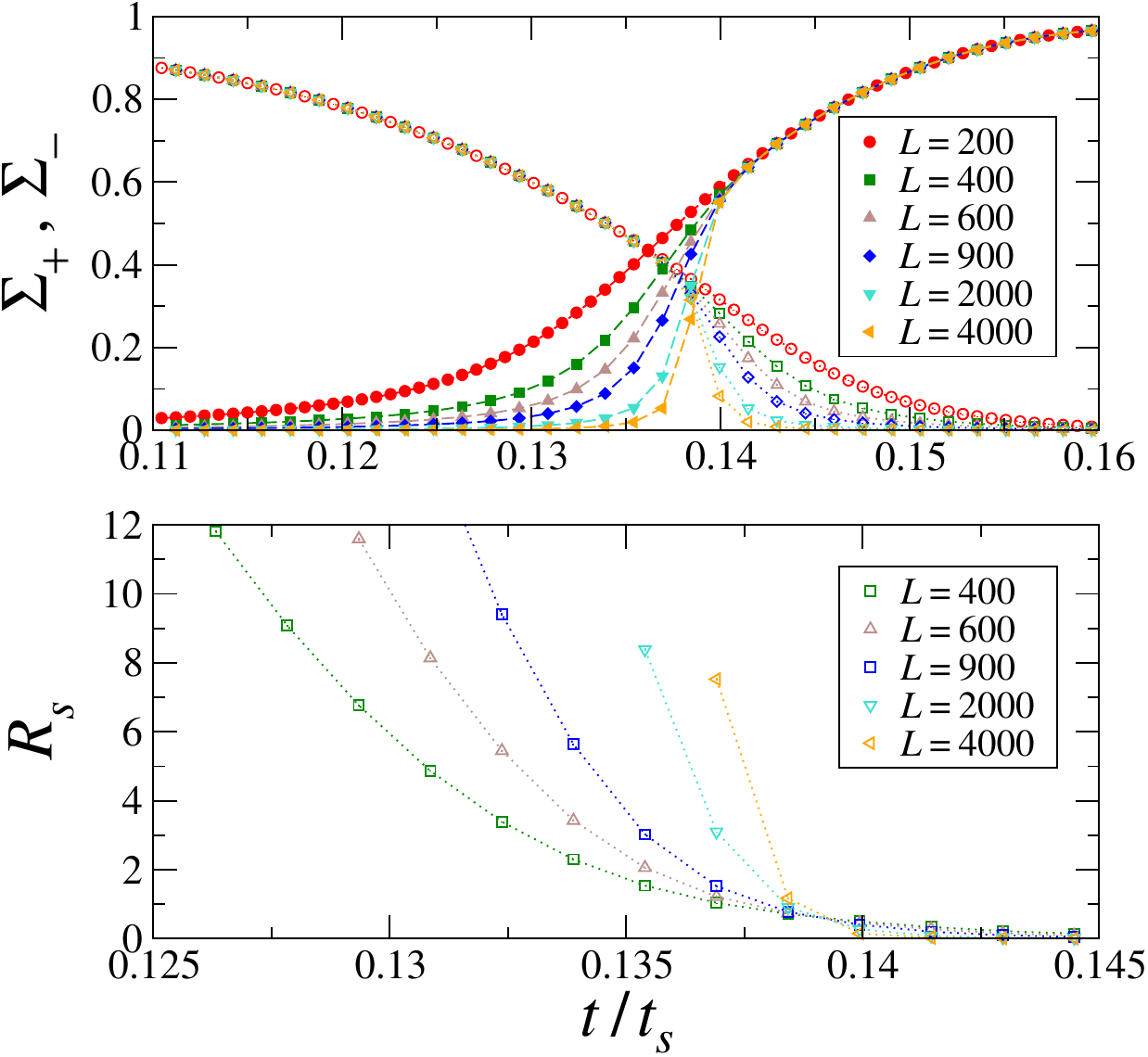}
  \caption{Results for $t_s = 5288$ versus $t/t_s$.
    Top: $\Sigma_+$ (filled symbols) and $\Sigma_-$ (empty symbols)
    for several values of $L$, for the adiabatic dynamics.
    Bottom: the ratio $R_s$ for several values of $L$. 
    Lines connecting the data points are only meant to guide the eye.}
  \label{Datavst_5288}
\end{figure}
%%%%%%%%%%%%%%%%%%%%%%%%%%%%%%%%%%%%%%%%%%%%%%%%%%%%%%%%%%%%%%%%%%%%%%%

The analysis of the spin clusters along the KZ protocol with slow
variations of $h$ [cf.~Eq.~\eqref{kzht}] shows behaviors analogous to
those observed in the case of sudden quenches.  In
Fig.~\ref{Datavst_5288} (top), we report the rescaled cluster sizes
$\Sigma_+$ and $\Sigma_-$ as a function of $t/t_s$ for $t_s = 5288$
and several values of $L$.  As the size increases, $\Sigma_+$ shows a
sharp change for $h(t)\approx 0.14$, corresponding to $t\approx
740$. For small values of $t$ none of the positive-magnetization
clusters fills a finite volume fraction of the lattice. On the other
hand, a fraction of the volume belongs to a single
negatively-magnetized percolating cluster. For $h(t)\gtrsim 0.14$
the opposite occurs, with a single positively-magnetized percolating
cluster coexisting with small negatively-magnetized domains. The
transition from the small-$t$ to the large-$t$ regime occurs abruptly
(for large systems) at a dynamical transition time $t_c$. As in the
quench case, with increasint $t$ positive-spin clusters aggregate very
rapidly close to $t_c$, forming a single large cluster and
partitioning the negatively-magnetized cluster that was filling the
space for $t < t_c$.

%%%%%%%%%%%%%%%%%%%%%%%%%%%%%%%%%%%%%%%%%%%%%%%%%%%%%%%%%%%%%%%%%%%%%%%
\begin{table}[t]
  \caption{For different values of $t_s$ we report the estimates of $t_c$,
    the corresponding values $t_c/t_s$ and $\sigma_c = (\ln t_c)^2
    t_c/t_s$, and the estimates of the exponent $w$. }
  \label{table-results-KZ}
  \begin{tabular}{llccc}
    \hline\hline
    $t_s$  & $t_c$ & $t_c/t_s$ & $\sigma_c$ &  $w$ \\
    \hline
    5288   &  734(4)   & 0.1388(8) & 6.04(4) & 0.71(2) \\
    10170  &  1200(5)  & 0.1180(5) & 5.93(3) & 0.71(2) \\
    20339  &  2036(5)  & 0.1001(2) & 5.81(2) & 0.68(2) \\
    42306  &  3589(5)  & 0.0848(1) & 5.68(1) & 0.65(2) \\
    \hline\hline
  \end{tabular}
\end{table}
%%%%%%%%%%%%%%%%%%%%%%%%%%%%%%%%%%%%%%%%%%%%%%%%%%%%%%%%%%%%%%%%%%%%%%%

%%%%%%%%%%%%%%%%%%%%%%%%%%%%%%%%%%%%%%%%%%%%%%%%%%%%%%%%%%%%%%%%%%%%%%%
\begin{figure}[t]
  \includegraphics[width=0.9\columnwidth]{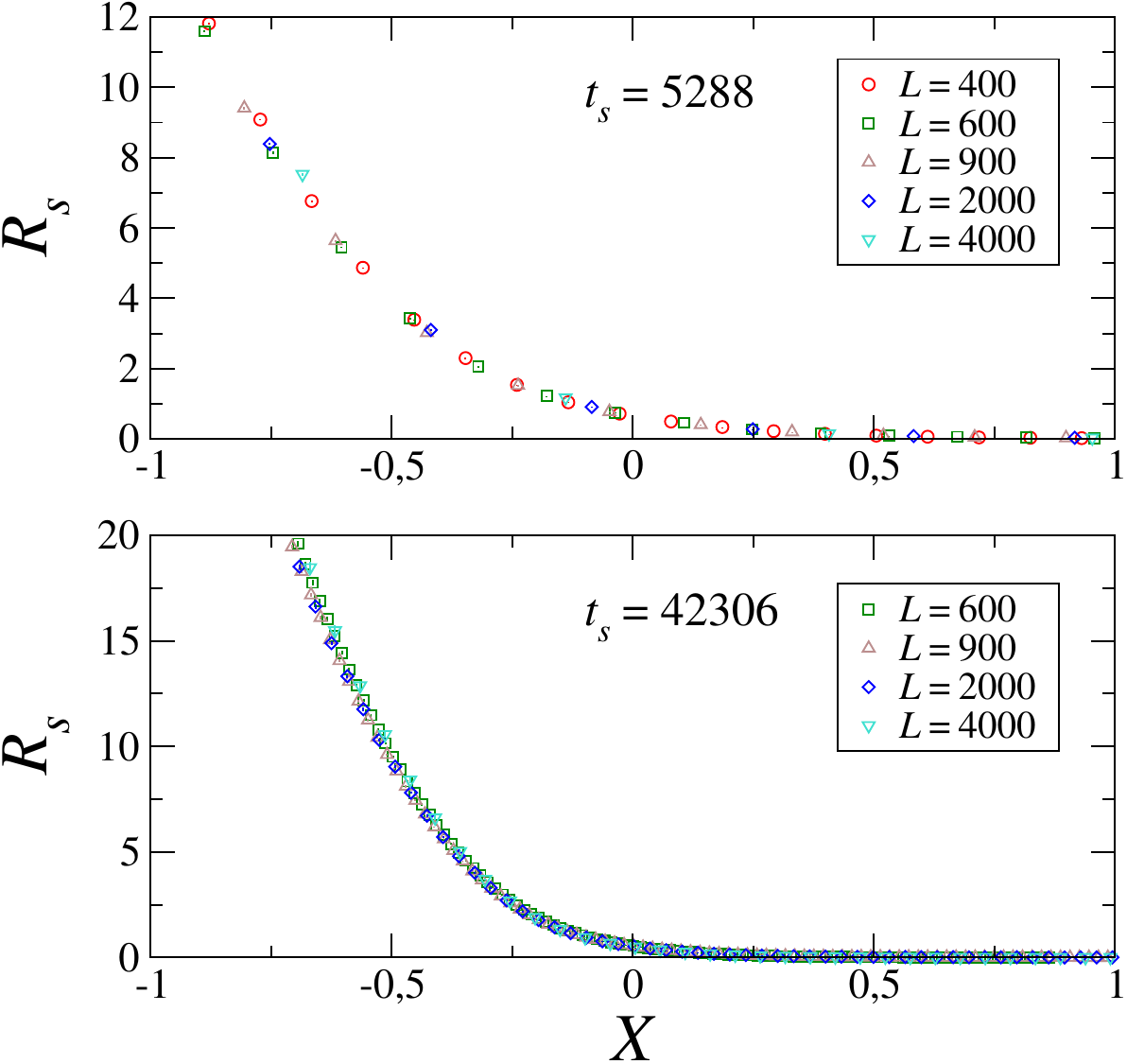}
  \caption{Plot of $R_s$ for several values of $L$ as a function of
    $X = (t - t_c) L^w/t_s$, for $t_s = 5288$ (top) and
    $t_s = 42306$ (bottom).  We use the values of $t_c$ and $w$
    reported in Table~\protect\ref{table-results-KZ}.}
  \label{RvsX-KZ}
\end{figure}
%%%%%%%%%%%%%%%%%%%%%%%%%%%%%%%%%%%%%%%%%%%%%%%%%%%%%%%%%%%%%%%%%%%%%%%

Figure~\ref{Datavst_5288} (bottom) shows data for the ratio
$R_s(t,t_s,L)$ of the sizes of largest negative and positive clusters
[cf.~Eq.~\eqref{drddef}], which appear to diverge for $h(t)\lesssim
0.14$ and to decrease towards zero in the opposite limit. Therefore,
$R_s(t,t_s,L)$ scales as in Eq.~\eqref{Ansatz-FSS}.  Straightforward
fits to the Ansatz~\eqref{Ansatz-FSS}---we approximate the function
${\cal F}_R(X)$ with a low-order polynomial---lead to the estimates
reported in Table~\ref{table-results-KZ}. Note that, as in the quench
case, the exponent $w$ depends on $t_s$ and decreases in the limit
$t_s\to\infty$. The resulting scaling behavior is quite good, as
reported in Fig.~\ref{RvsX-KZ}.

%%%%%%%%%%%%%%%%%%%%%%%%%%%%%%%%%%%%%%%%%%%%%%%%%%%%%%%%%%%%%%%%%%%%%%%
\begin{figure}[tbp]
  \includegraphics[width=0.9\columnwidth]{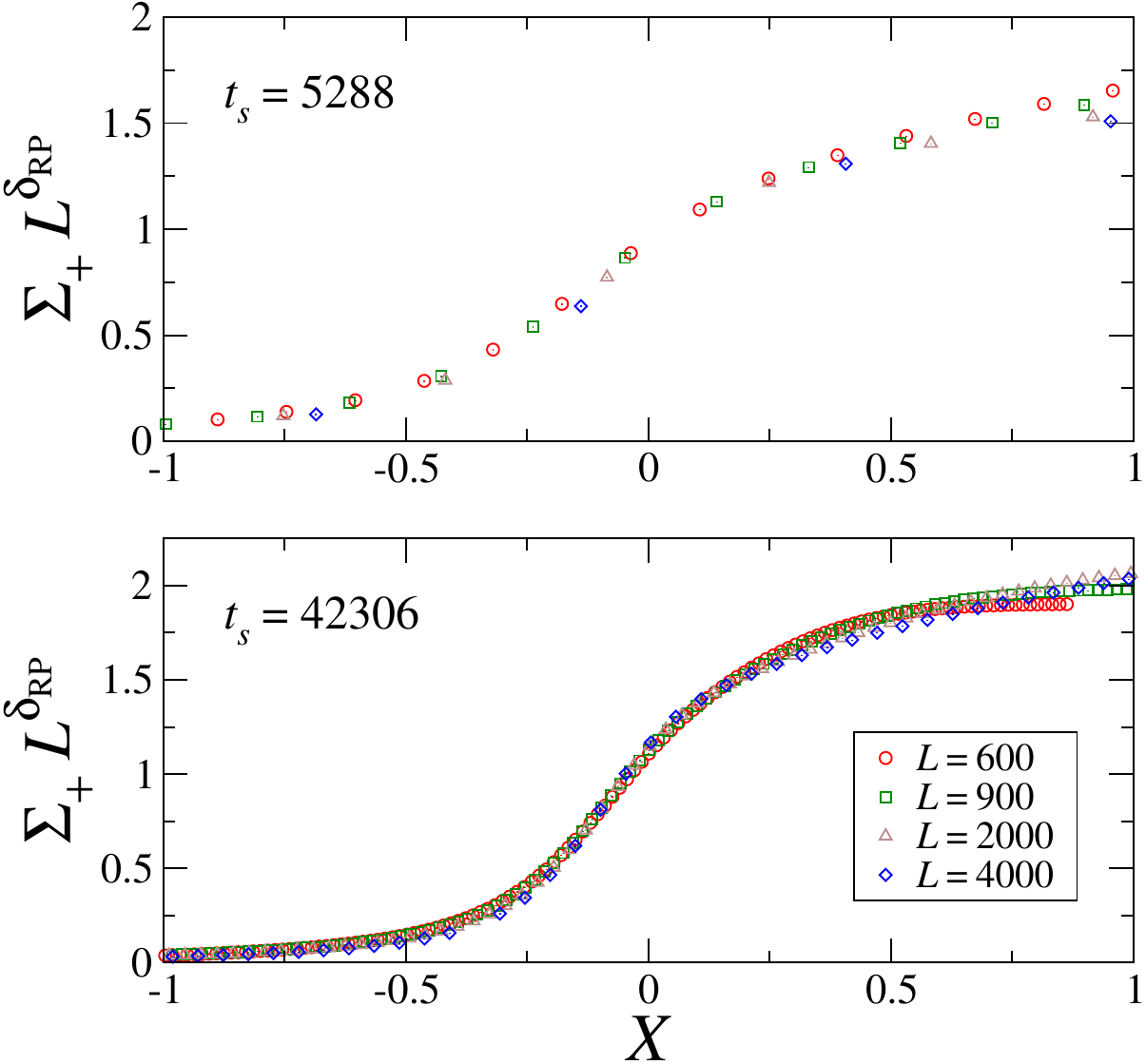}
  \caption{Plot of $\Sigma_+ L^{\delta_{\rm RP}}$ 
    versus $X = (t-t_c) L^w/t_s$, for two values of $t_s$.
    We use the estimates of $t_c$ and $w$ reported in
    Table~\ref{table-results-KZ} and $\delta_{\rm RP} = 5/48$.
    Results are for $t_s = 5288$ (top) and $t_s = 42306$ (bottom).}
  \label{Spiuscal}
\end{figure}
%%%%%%%%%%%%%%%%%%%%%%%%%%%%%%%%%%%%%%%%%%%%%%%%%%%%%%%%%%%%%%%%%%%%%%%

%%%%%%%%%%%%%%%%%%%%%%%%%%%%%%%%%%%%%%%%%%%%%%%%%%%%%%%%%%%%%%%%%%%%%%%
\begin{figure}[tbp]
  \includegraphics[width=0.9\columnwidth]{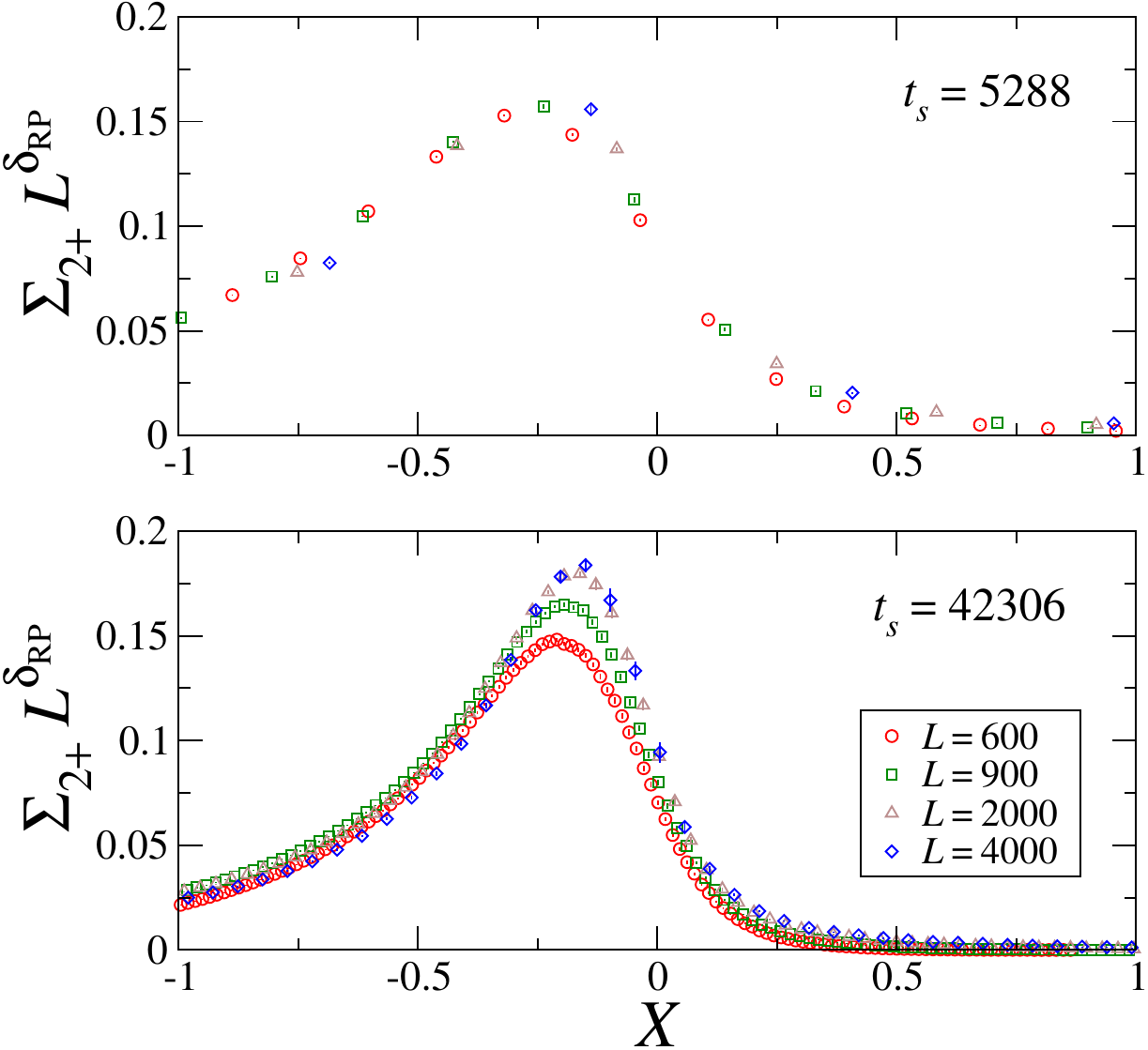}
  \caption{Plot of $\Sigma_{2+} L^{\delta_{\rm RP}}$ versus $X=
    (t-t_c) L^w/t_s$, for two values of $t_s$. We use the estimates of
    $t_c$ and $w$ reported in Table~\ref{table-results-KZ} and
    $\delta_{\rm RP} = 5/48$. Results are for $t_s = 5288$ (top)
    and $t_s = 42306$ (bottom).}
  \label{Spiu2scal}
\end{figure}
%%%%%%%%%%%%%%%%%%%%%%%%%%%%%%%%%%%%%%%%%%%%%%%%%%%%%%%%%%%%%%%%%%%%%%%

Finally, we determine the fractal dimension of the clusters.  In the
analysis of the quench case, we showed that the data for $\Sigma_+$
and $\Sigma_-$ were consistent with the RP fractal dimension $d_{\rm
  RP} = 2 -\delta_{\rm RP}$, $\delta_{\rm RP} = 5/48 \approx
0.104$. The same result holds for the KZ case. Indeed, plotting
$\Sigma_+ L^{\delta_{\rm RP}}$ versus $X$ as in Fig.~\ref{Spiuscal},
using the values of $t_c$ and $w$ reported in
Table~\ref{table-results-KZ}, data nicely fall onto a single curve,
confirming the scaling Ansatz~\ref{Dpiu-scaling}. Interestingly,
the size of the second-largest cluster satisfies the same scaling
Ansatz, as shown inFig.~\ref{Spiu2scal}; the same holds
for the third-largest cluster size (not shown).

\subsection{Scaling of the magnetization}

%%%%%%%%%%%%%%%%%%%%%%%%%%%%%%%%%%%%%%%%%%%%%%%%%%%%%%%%%%%%%%%%%%%%%%%
\begin{figure}[tbp]
  \includegraphics[width=0.9\columnwidth]{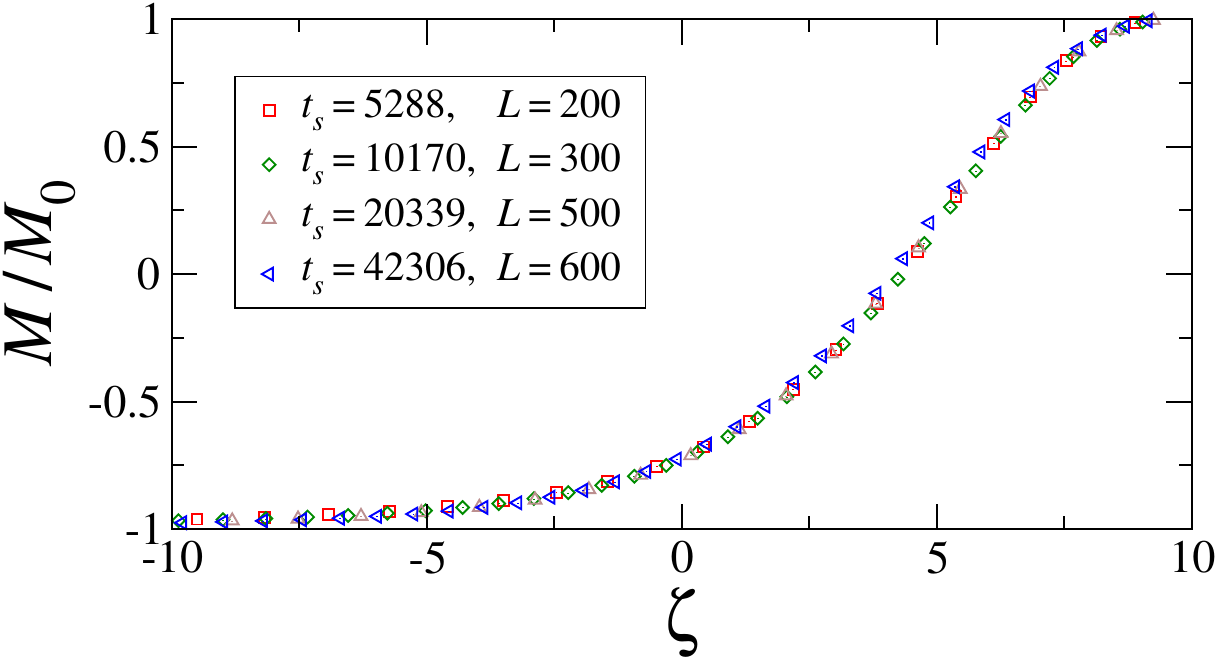}
  \caption{The rescaled magnetization $M/M_0$ in the infinite-volume
    limit, as a function of $\zeta = (\sigma-\sigma_*) \, h(t)^{-\theta}$.
    Here $\sigma_* = 4.75$ and $\theta = 0.70$.}
  \label{rescKZ}
\end{figure}
%%%%%%%%%%%%%%%%%%%%%%%%%%%%%%%%%%%%%%%%%%%%%%%%%%%%%%%%%%%%%%%%%%%%%%%

We now discuss the infinite-volume magnetization $M_{\infty}(t,t_s)$,
obtained by considering the limit $L\to\infty$ at fixed $t_s$.  The
arguments presented in Sec.~\ref{sec2.B1} also lead to the hypothesis
that the time-dependent magnetization can be parametrized in terms of
the scaling variable
\begin{equation}
   \sigma = (\ln t)^2 h(t) = (\ln t)^2{t\over t_s} .
\end{equation}
Correspondingly, as numerically verified in Ref.~\cite{PV-26}, for 
$t_s\to \infty$, 
the magnetization obeys the scaling relation~\eqref{Mscaling}.
When plotted in the terms of $\sigma$, the finite-$t_s$ curves 
intersect at a particular value $\sigma=\sigma_*$, around which
the data can be parametrized as
\begin{equation}
  M_{\infty}(t,t_s) \approx \widehat{\cal M}(\bar{\zeta}), \qquad
  \bar{\zeta} = (\sigma - \sigma_*) t_s^{\overline{\theta}} \,.
  \label{scalingKZ}
\end{equation}
For $\beta = 1.2\beta_c$, the analysis of the numerical data along the
KZ protocol provided the estimates~\cite{PV-26}
\begin{equation}
\sigma_* = 4.59(9), \qquad {\overline\theta} = 0.145(20).
\end{equation}
We now show that the data are also consistent with the Ansatz used in
the quench case,
\begin{equation}
  M_{\infty}(t,t_s)  = \widetilde{\cal M}(\zeta),
  \qquad 
  \zeta = (\sigma - \sigma_*) h(t)^{-{\theta}}.
  \label{scalingm-quench}
\end{equation}
In fact, fitting our magnetization data to
Eq.~\eqref{scalingm-quench}, we obtain a good scaling behavior, as
shown in Fig.~\ref{rescKZ}, by fixing $\sigma_* = 4.75(15)$,
consistent with the estimate $\sigma_* = 4.59(9)$ of
Ref.~\cite{PV-26}, and $\theta = 0.70(10)$.  The exponent $\theta$
obtained here is consistent with the result obtained for the quench
case, $\theta = 0.64(4)$, further indicating that the underlying
physics is the same in the two cases.

Note that the scaling behaviors~\eqref{scalingKZ}
and~\eqref{scalingm-quench} are not equivalent, although they can be
hardly distinguished numerically.  Indeed, since $h(t) = t/t_s =
\sigma (\ln t)^{-2}$, for fixed $\sigma$ and $t_s\to \infty$, we have
$t \approx \sigma t_s (\ln \sigma t_s)^{-2}$, with corrections of
$O(\ln\ln \sigma t_s/\ln \sigma t_s)$.  Therefore, we can
asymptotically rewrite $\zeta$ as
\begin{equation}
  \zeta \sim (\sigma - \sigma_*) (\ln \sigma_* t_s)^{2\theta},
  \label{zetatsstar}
\end{equation}  
differing from $\overline{\zeta}$ by the fact that $(\ln
t_s)^{2\theta}$ is replaced by $t_s^{\overline\theta}$.  The two
scaling variables are numerically equivalent, since a logarithmic
term cannot be numerically distinguished from a power term with a
small exponent ($\bar{\theta}\approx 0.14$ is indeed small).

The value $\sigma_*$ corresponds to a time $t_*$ defined by
requiring $\sigma_* = (\ln t_*)^2 \,t_*/t_s$. Thus,
for large $t_s$ we have
\begin{equation}
  t_* \approx  {\sigma_* t_s \over (\ln \sigma_* t_s)^2} ,
  \label{tstarts}  
\end{equation} 
showing that $h(t_*)$ vanishes for $t_s\to \infty$ as 
$(\ln \sigma_* t_s)^{-2}$.

In Table~\ref{table-results-KZ} we also report the value $\sigma_c$
of $\sigma$ corresponding to the finite-time transition point $t_c$,
i.e., $\sigma_c = (\ln t_c)^2 t_c/t_s$.  In the quench case, we showed
that $\sigma_c$ converges to $\sigma_*$ with corrections that scale as
$h^\theta$. Using the results of Table ~\ref{table-results-KZ} to
compute
\begin{equation}
  \zeta_c = (\sigma_c - \sigma_*) h(t_c)^{-{\theta}},
  \label{zetacdef}
\end{equation}
we obtain $5.15(13)$, $5.27(11)$, $5.31(7)$, $5.25(4)$ for $t_s = 5288$,
$10170$, $20339$, $42306$, respectively. Errors take only into account
the error on $t_c$, but not that on $\sigma_*$ and $\theta$.
Indeed, such uncertainties affect the estimates of $\zeta_c$ for different
values of $t_s$ in the same way and thus they are not relevant to
verify that $\zeta_c$ is independent of $t_s$. These results imply
\begin{equation}
  t_c = t_* (1 + e h^\theta),\qquad 
  e = {\zeta_c\over 2 \sigma_*} \approx 0.6.
  \label{tctstare}
\end{equation}
Thus, in the small-$h$ limit, the dynamic percolation time $t_c$, at
which the abrupt aggregation of the positive-spin clusters and the
fragmentation of the percolating negative-spin cluster characterizing
the configurations for $t < t_c$ are observed, approaches the time
$t_*$ in Eq.~\eqref{tstarts}.

\section{Conclusions}
\label{conclu}

We consider Ising-like systems evolving under a purely relaxational
dynamics without conservation laws, and investigate their out-of-equilibrium
behavior following a quench of the external magnetic field $h$ across
the low-temperature FOT line.  Specifically, we focus on
the paradigmatic square-lattice nearest-neighbor Ising model evolving
under the heat-bath or the Metropolis relaxational dynamics, and study
how negatively-magnetized configurations, thermalized at negative
values of the magnetic field $h$, evolve after switching to a positive
field at fixed, sufficiently low temperature.

Our results reveal the emergence of a dynamic percolation transition
at a finite post-quench critical time $t_c$, for finite
(yet sufficiently small) values of $h$, say $h\lesssim 0.1$.  This
transition is marked by the simultaneous percolation of the largest
positive cluster and inverse percolation of the largest negative
one, signaling the transition from the metastable
negative-magnetization phase to the stable positive one.  For times
$t$ close to $t_c$, the size of the largest clusters satisfies the FSS
relation~\eqref{Ansatz-FSS}, characterized by the fractal dimensions
$d_\pm$ and the critical exponent $w$.  The fractal dimensions
$d_\pm$ turn out to be equal to the RP value $d_{\rm RP} = 91/48$, and
are apparently independent of $h$.  In contrast, the exponent
$w$, which controls the approach to criticality, differs from its RP
value $w_{\rm RP}=3/4$ and decreases as $h$ is reduced, apparently as
$w\propto h^\theta$ with $\theta\approx 0.6$.

From a renormalization-group point of view, this behavior may be
interpreted as arising from a line of $h$-dependent dynamic fixed
points that share the same metric (magnetic, in the equivalent spin
representation) critical exponents, but differ in the scaling behavior
governing the approach to criticality.  A similar phenomenon is
observed in the equilibrium critical behavior of the 2D classical
Ashkin-Teller (AT) model and of its one-dimensional quantum
counterpart (see, e.g.,
Refs.~\cite{AT-43,WL-74,Wu-77,Baxter-82,YHK-94,YK-95,RV-26}).  The AT
model shows a line of critical transitions, along which the exponent
$\eta$ (corresponding to the fractal dimension in percolation) remains
constant, while the length-scale exponent $\nu$ varies.

The critical percolation time $t_c$ is related to the characteristic
time $t_*$ at which the system switches from the metastable negatively
magnetized phase to the positively magnetized one, and hence to the
spinodal-like behavior of the magnetization in the small-$h$
limit. Consequently, $t_c$ also exhibits an exponential dependence on
$h$, as in Eq.~\eqref{tcsca}, qualitatively resembling false-vacuum
decay pictures.  Our results suggest that the transition from the
metastable to the stable phase is not driven by the independent growth
of isolated droplets, but rather by their collective aggregation, with
droplets growing predominantly through merging and coalescence.

The percolative behavior observed along the post-quench dynamics of 2D
Ising systems across their FOTs is likely to extend to a broader class
of ferromagnetic Ising-like systems.  Percolation transitions may thus
also occur under alternative quenching
protocols~\cite{LP-90,PV-17,PV-24,PV-26,PRV-26,HSZL-24} and in higher
dimensions. For example, as discussed in Sec.~\ref{slowdyn}, analogous
features are observed when the system is slowly driven across the FOT
line according to KZ-like crossing protocols.

Note that some qualitative features of the dynamic percolation
behavior may depend on the lattice system, in particular on its
geometry and dimensions.  For example, in three-dimensional Ising
systems~\cite{BPV-26}, positive and negative clusters undergo distinct
percolation transitions, giving rise to an intermediate dynamical
phase in which percolating clusters of both signs coexist. Moreover,
we cannot exclude the possibility that the coincidence of the
percolation transitions of the positive and negative clusters observed
here is a consequence of the self-duality of the square lattice,
rather than a generic feature of the dynamics. If so, other 2D
lattices with different geometries may exhibit two distinct
percolation transitions, with an intermediate phase characterized
either by the coexistence of positive and negative percolating
clusters, as in three dimensions, or by their absence.
More generally, percolation transitions are only expected in systems
with discrete symmetries. Systems with continuous symmetries may
exhibit qualitatively different behaviors, due to the presence of
Goldstone modes~\cite{PV-16}. The resulting phenomenology may depend
on both the dimensionality and the number of low-temperature phases.
Another related issue is whether percolation phenomena also occur at
temperature-driven FOTs, such as those occurring in $q$-state Potts
models for $q\ge 3$. These further issues call for future
investigations.

We finally remark that the critical features of the out-of-equilibrium
percolation transitions observed across the FOTs of Ising systems
differ from those of the standard RP universality class.  Non-RP
finite-time transitions have also been reported in network models
subject to nonlocal and irreversible dynamics, with explosive percolation
(EP) transitions providing a notable
example~\cite{AGKSZ-14,ASS-09,Ziff-09,CKPKK-09,RF-09,Ziff-10,CDGM-10,DM-10,
  RF-10,CD-11,NLT-11,GCBSP-11,CK-11,LKP-11,RW-11,RW-12,NTG-12,CDGM-14,
  BGMK-14,DN-15,Boetal-16,HS-18,LWD-23,YL-24}. In those cases, however,
deviations from RP are essentially associated with the {\em ad hoc}
nonlocal nature of the underlying dynamics.  By contrast, the
percolation transitions considered here arise from entirely local
relaxational dynamics. Their deviation from standard RP behavior
stems from the intrinsically out-of-equilibrium nature of the dynamics
across FOTs. Therefore our results point to a distinct class of non-RP
finite-time percolative phenomena that can naturally emerge in physical
systems driven across FOTs by physically realizable local dynamics.
This contrasts with other examples of non-RP finite-time transitions
arising from nonlocal and irreversible {\em ad hoc} dynamical rules,
as in certain network-growth models exhibiting EP
transitions~\cite{ASS-09,Ziff-09,CKPKK-09,RF-09,Ziff-10,CDGM-10,DM-10,
  RF-10,CD-11,NLT-11,GCBSP-11,CK-11,LKP-11,RW-11,RW-12,NTG-12,CDGM-14,
  BGMK-14,DN-15,Boetal-16,HS-18,LWD-23,YL-24}.

\end{document}